\documentclass[10pt,aps,prb,twocolumn,superscriptaddress,showpacs]{revtex4-2}

\usepackage{graphicx}
\usepackage{amsmath}
\usepackage{amssymb}
\usepackage{hyperref}
\usepackage{balance}
\usepackage{braket}
\usepackage{xcolor}

\begin{document}

\title{Quasiparticle Dynamics in the 4d--4f Ising-like Double Perovskite Ba$_2$DyRuO$_6$ studied using Neutron Scattering and Machine-Learning Framework}

\author{G. Roy}
\email{gourabr22bs@rgipt.ac.in}
\affiliation{Rajiv Gandhi Institute of Petroleum Technology, Jais, Amethi, 229304, India}

\author{E. Kushwaha}
\affiliation{Rajiv Gandhi Institute of Petroleum Technology, Jais, Amethi, 229304, India}

\author{M. Kumar}
\affiliation{Rajiv Gandhi Institute of Petroleum Technology, Jais, Amethi, 229304, India}

\author{S. Ghosh}
\affiliation{Rajiv Gandhi Institute of Petroleum Technology, Jais, Amethi, 229304, India}

\author{F. Orlandi}
\affiliation{ISIS Neutron and Muon Source, STFC, Rutherford Appleton Laboratory, Harwell campus, Didcot, Oxfordshire OX11-0QX, United Kingdom}

\author{M. D. Le}
\affiliation{ISIS Neutron and Muon Source, STFC, Rutherford Appleton Laboratory, Harwell campus, Didcot, Oxfordshire OX11-0QX, United Kingdom}

\author{M. B. Stone}
\affiliation{Neutron Scattering Division, Oak Ridge National Laboratory, Oak Ridge, Tennessee 37831, USA}

\author{J. Sannigrahi}
\affiliation{School of Physical Sciences, Indian Institute of Technology Goa, Farmagudi, Goa 403401, India}

\author{D. T. Adroja}
\affiliation{ISIS Neutron and Muon Source, STFC, Rutherford Appleton Laboratory, Harwell campus, Didcot, Oxfordshire OX11-0QX, United Kingdom}
\affiliation{Highly Correlated Matter Research Group, Physics Department,University of Johannesburg, Auckland Park 2006, South Africa}

\author{T. Basu}
\email{tathamay.basu@rgipt.ac.in}
\affiliation{Rajiv Gandhi Institute of Petroleum Technology, Jais, Amethi, 229304, India}

%\date{\today}

\begin{abstract}
Double perovskites containing 4d--4f interactions provide a platform to study complex magnetic phenomena in correlated systems. Here, we investigate the magnetic ground state and quasiparticle excitations of the fascinating double perovskite system, Ba$_2$DyRuO$_6$, through Time of flight (TOF) neutron diffraction, inelastic neutron scattering (INS), and theoretical modelling. The compound Ba$_2$DyRuO$_6$ is reported to exhibit a single magnetic transition, in sharp contrast to most of the other rare-earth (R) members in this family, A$_2$RRuO$_6$ (A = Ca/Sr/Ba), which typically show magnetic ordering of the Ru ions, followed by R-ion ordering. Our neutron diffraction results confirm that long-range antiferromagnetic order emerges at $T_\mathrm{N} \approx 47$~K, primarily driven by 4d--4f Ru$^{5+}$--Dy$^{3+}$ exchange interactions, where both Dy and Ru moments contribute to the ordered state. The ordered ground state is a collinear antiferromagnet with Ising character, carrying ordered moments of $\mu_{\mathrm{Ru}} = 1.6(1)~\mu_\mathrm{B}$ and $\mu_{\mathrm{Dy}} = 5.1(1)~\mu_\mathrm{B}$ at 1.5~K. Low-temperature INS reveals well-defined magnon excitations below 10~meV. SpinW modelling of the INS spectra evidences complex exchange interactions and the presence of magnetic anisotropy, which governs the Ising ground state and accounts for the observed magnon spectrum. Combined INS and Raman spectroscopy reveal crystal-electric-field (CEF) excitations of Dy$^{3+}$ at 46.5 and 71.8~meV in the paramagnetic region. The observed CEF levels are modelled through point-charge CEF calculations, consistent with the $O_h$ site symmetry of Dy$^{3+}$. A complementary machine-learning approach is used to calculate and analyse the phonon spectrum, enabling a direct comparison with the INS data. These combined experimental and theoretical results provide a comprehensive understanding of the origin of phonon and magnon quasiparticle excitations and their influence on the crystal structure and ground-state magnetism of Ba$_2$DyRuO$_6$.

\end{abstract}

 \maketitle

\section{Introduction}

The strong $d$--$f$ interactions in rare-earth (R) transition-metal (M) oxides provide valuable insights into exotic magnetic and electronic properties\cite{ref1,ref2,ref3,ref4,ref5,ref6,ref7,ref8,ref9}. While $3d$--$4f$ systems have been extensively studied \cite{ref1,ref5,ref6,ref7,ref8, ref10}, materials incorporating both rare-earth (R) and $4d/5d$ transition metal (TM) ions have received comparatively less attention. In particular, the $4d$--$4f$ coupling benefits from enhanced hybridization, strong spin-orbit coupling (SOC), and crystal electric field (CEF) effects of the $4d$ orbitals combined with rare-earth anisotropy of $f$-orbitals, leading to unconventional magnetic ground states\cite{ref2,ref11,ref12}. The interplay between all these key factors enables the design of tunable magnetic materials, advancing research in quantum magnetism. For example, the $4d$--$4f$ pyrochlore systems, R$_2$Ru$_2$O$_7$ (R = Y, Pr, Nd, Gd, Dy, Ho, etc.), exhibit fascinating magnetic ground states at different temperatures, such as spin-frustrated states and spin-ice behavior, with different magnetic ground states for different R-ions~\cite{ref7,ref15,ref16,ref17}. Recent studies on the 6H-perovskite Ba$_3$RRu$_2$O$_9$ document versatile physical properties and intriguing magnetic ground states for different rare-earth members, such as cooperative magnetic ordering of R and Ru moments~\cite{ref2,ref11,ref12,ref13}, spin-driven ferroelectricity for the Ho member~\cite{ref11,ref18,ref19}, an unusual $+4$ valence state for Tb~\cite{ref12}, and most surprisingly, ferromagnetic ordering of Nd in the Ba$_3$NdRu$_2$O$_9$ compound~\cite{ref20}. The versatile behavior could result from differing degrees of $4d$--$4f$ hybridization and lattice distortion, both of which play decisive roles in determining the exchange interactions. 

Double perovskites with larger $d$ orbitals ($4d/5d$) have gained additional attention, as they exhibit a variety of rich phenomena such as spin frustration, high-temperature ordering via double exchange interactions, field-induced magnetism, and magnetoresistance~\cite{ref14,ref21,ref22,ref23,ref24,ref25,ref26,ref27}. In particular, the A$_2$LnRuO$_6$ series (A = alkali metal; Ln= Y, La, Lu (nonmagnetic) or rare--earth ions Ce--Yb (magnetic)) shows exotic ground states arising from spin frustration and complex magnetic exchange interactions driven by A-site and Ln-site cation variations~\cite{ref14,ref28,ref29,ref30,ref31,ref32,ref33,ref34,ref35,ref36,ref37}. For instance, Sr$_2$LnRuO$_6$ (Ln = Dy, Ho, Er) crystallizes in the monoclinic space group $P2_1/n$ with two phase transitions, exhibiting exchange-bias properties that may arise due to the Dzyaloshinskii--Moriya (DM) interaction in such low-symmetry crystal structures~\cite{ref32}. In contrast, substituting nonmagnetic Y$^{3+}$ for Ln, while retaining the same space group, reveals planar AFM correlations above 24~K and a partially ordered state between 24--32~K, attributed to geometrical frustration inherent to the lattice~\cite{ref14}. Moreover, in such double perovskites, the nonmagnetic alkaline-earth metal (A = Ba, Sr, Ca) plays a pivotal role in the structural stability and electronic environment, which in turn vastly modifies the magnetic exchange interactions, yielding a completely different magnetic ground state for different A-ion. For example, most Sr-based double perovskites discussed above, such as Sr$_2$RRuO$_6$ (R = Y, Dy, Ho, Er, Tb), crystallize in the monoclinic space group P$2_1$/n~\cite{ref14,ref32,ref33,ref34,ref35}. When Sr is replaced by the larger Ba atom, such as, Ba$_2$DyRuO$_6$~\cite{ref29}, Ba$_2$HoRuO$_6$~\cite{ref30}, Ba$_2$YRuO$_6$~\cite{ref36}, the system crystallises in a highly symmetric Fm-$3$m cubic structure, with the only exception of Ba$_2$PrRuO$_6$~\cite{ref31}, which crystalline in a monoclinic space group $P2_1/n$. However, in all these cubic systems the magnetic ground-state remains poorly understood. Given their wide range of technological applications, it is important to explore how changing each element--either A (nonmagnetic) or B (magnetic)--can significantly alter the crystal structure, and how the associated symmetries strongly influence the ground-state magnetic properties in this diverse group of materials.

\begin{figure}[!ht]
    \centering
    \includegraphics[width=.5\textwidth]{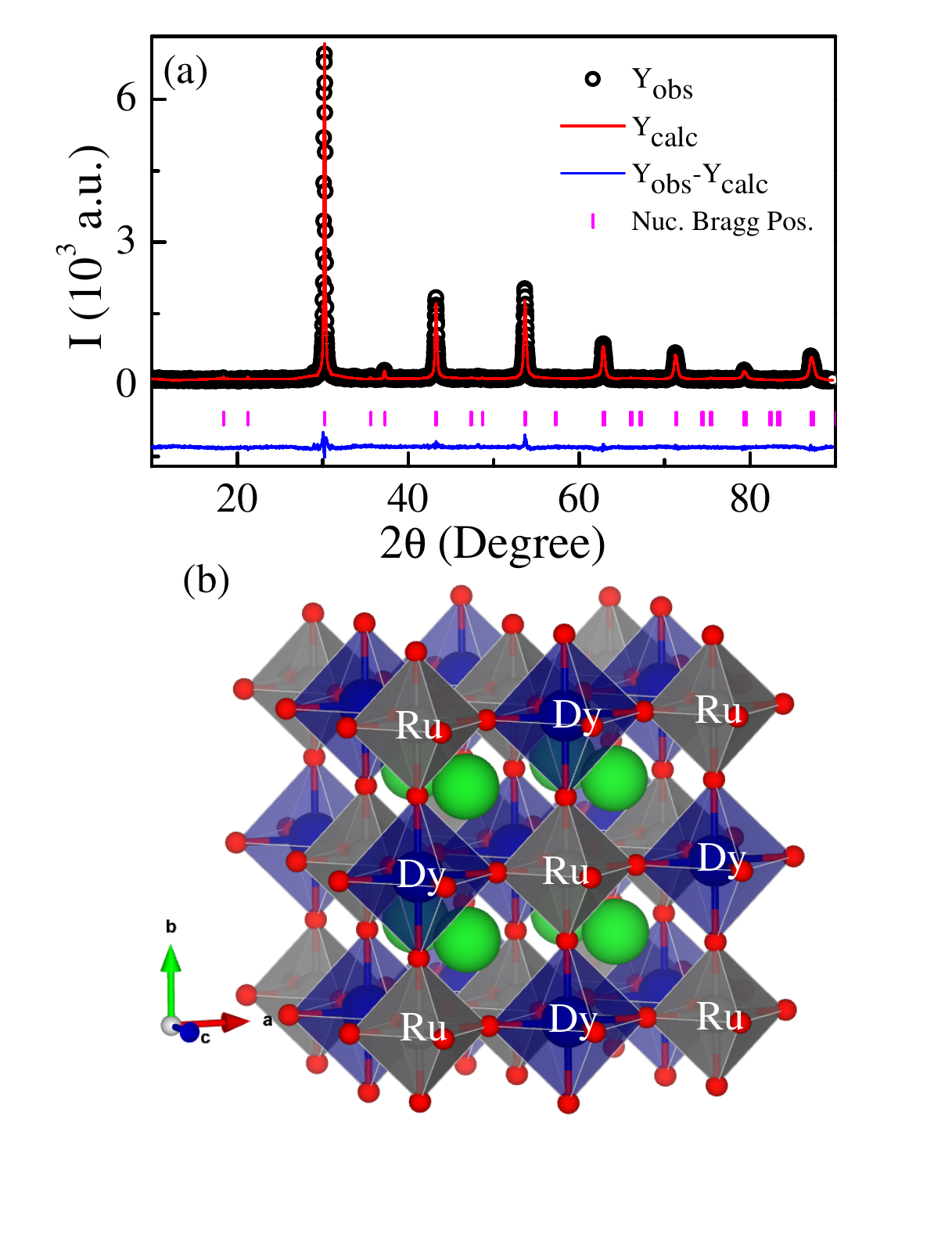} 
    \caption{ (a) Rietveld refinement of the X-ray diffraction (XRD) pattern of Ba$_2$DyRuO$_6$ measured at room temperature, with intensity (I) plotted as a function of 2$\theta$ (°). Open black circles represent the observed data, the solid red line corresponds to the calculated pattern, vertical pink ticks indicate Bragg reflection positions, and the blue line at the bottom shows the difference between observed and calculated intensities.
    (b) Crystal structure of Ba$_2$DyRuO$_6$ crystallizing in the cubic Fm-$3$m space group. Red spheres represent oxygen sites and green spheres represent barium sites.
}
    \label{fig:fig1}
\end{figure}

 Ba$_2$DyRuO$_6$ crystallises in a cubic structure with space group Fm-$3$m . The experimentally determined effective paramagnetic moment, $\mu_{\mathrm{eff}} = 11.08\,\mu_{\mathrm{B}}$, is close to the expected value of $11.28\,\mu_{\mathrm{B}}$, consistent with the presence of Dy$^{3+}$ and Ru$^{5+}$ ions in Ba$_2$DyRuO$_6$ ~\cite{ref29}. It exhibits a single magnetic ordering transition at 47~K, in sharp contrast to the previously discussed $R$ ions in the same family, where two distinct ordering temperatures were observed~\cite{ref29}. The ground states of Ru and Dy in this compound remain unexplored. Here, we investigate this system using time-of-flight neutron powder diffraction (NPD), inelastic neutron scattering (INS), and modelling  the magnetic (or spin wave) excitations using \textsc{SpinW} simulations~\cite{ref38}. Additionally, theoretical simulations and machine-learning methods were employed to understand the crystal electric field (CEF) and Phonon excitations in this system. By examining these results collectively, we have estimated the nearest- and next-nearest-neighbour exchange interactions and magnetic anisotropy, modelling the INS spectra with \textsc{SpinW} simulations. Such a comprehensive investigation is scarce in the literature for this entire series.

\section{Experimental Details}

The polycrystalline sample Ba$_2$DyRuO$_6$ is prepared by solid-state reaction by mixing and grinding BaCO$_3$, Dy$_2$O$_3$, and RuO$_2$ in a mortar and pestle in the proper ratio ~\cite{ref29}. The sample is initially heated at $960^{\circ}$C for 24 hours in the first step. The final sintering is performed on the pelletized sample at $1140^{\circ}$C for 24 hours with several intermediate mixing and grinding of the sample to obtain a uniform homogenised sample. Then, powder X-ray diffraction (XRD) patterns were recorded at room temperature using Cu~K\(\alpha\) radiation (\(\lambda = 1.5406~\text{\AA}\)) on a PANalytical Empyrean diffractometer. The room-temperature Rietveld refinement (FIG.~1(a)), performed using the FullProf Suite~\cite{ref39}, confirms that the sample forms in the desired phase, crystallizing in the cubic Fm-$3$m space group (FIG.~1(b)), consistent with earlier reports~\cite{ref29}. The Dy$^{3+}$ and Ru$^{5+}$ ions occupy the Wyckoff positions $4b$ and $4a$, respectively, in a 1:1 ratio. The lattice parameters obtained from the refinement are $a = b = c = 8.345(2)\ \text{\AA}$ and $\alpha = \beta = \gamma = 90^\circ$, consistent with the earlier report ~\cite{ref29}. Neutron Powder Diffraction (NPD) is performed using the high-flux and high-resolution Time-of-Flight WISH Diffractometer at ISIS, UK~\cite{ref40}. The 4.5\,g sample was packed in a sealed vanadium annular can for the NPD measurements. Initially, room-temperature data were taken, and then the sample was cooled using a liquid He cryostat. The low-temperature NPD data were collected during heating from $1.5$ K to $250$ K. The refinement and analysis of NPD data were carried out using the FullProf~\cite{ref39}, BASIREPS~\cite{ref41}, VESTA~\cite{ref42} software packages and Bilbao crystallographic server for magnetic space group~\cite{ref43}. Inelastic neutron scattering study was conducted on the MARI spectrometer at ISIS, UK, using a $3$ g sample in an aluminium canister~\cite{ref44}. The incident neutron energy was set at $E_i = 180$ meV (with additional measurements at $100$, $29.7$, $22.9$, and $11.7$ meV) in Repetition Rate Multiplication(RRM) mode, with a Gd-Fermi chopper operating at $400$ Hz. Data were collected at temperatures of $5$ K, $30$ K, and $100$ K. The analysis of INS data was performed using the Mantid and Dave software packages~\cite{ref45,ref46}. Further theoretical modelling was carried out using SpinW, a MATLAB-based library~\cite{ref38}. SpinW numerically computes and visualises the spin-wave dispersion by employing linear spin wave theory, incorporating the spin Hamiltonian, magnetic structure, and structural and magnetic twinning effects. The Raman spectrum was recorded at room temperature using a $532$ nm excitation laser, a $1200$ grooves/mm grating, $100$ accumulations, and a spectral range of $200–2000$ cm\(^{-1}\) at $1$ cm\(^{-1}\) resolution. Phonon calculations were carried out using the INSPIRED software~\cite{ref47}, which employs pre-trained machine learning force fields (MLFFs) derived from the MatterSim modeling package. The optimized structure and momentum-dependent phonon modes were subsequently obtained~\cite{ref48}.

\section{Results and Discussions}

\subsection{Neutron Powder Diffraction}  

Neutron powder diffraction (NPD) measurements were performed on Ba$_2$DyRuO$_6$ at several temperatures down to $1.5$~K to determine the ground state magnetic structure ~\cite{ref49}. The Rietveld refinements at $300$ K in the low-$Q$ range ($0.65 \le Q \le 2~\mathrm{\AA^{-1}}$) and the high-$Q$ range ($2 \le Q \le 12~\mathrm{\AA^{-1}}$) are shown in FIGs.~3(a) and 3(b), while the corresponding refinements at $1.5$ K are shown in FIGs.~3(c) and 3(d). Initially, the $300$ K data were refined to determine the crystal structure, followed by the $1.5$ K data to explore the magnetic structure. The cell parameters, atomic positions, thermal parameters, and occupancies of the nuclear structure obtained from the Rietveld refinement are listed in TABLE~I for $1.5$ K and $300$ K. The neutron absorption of Dy was properly accounted for during the refinement by using an empirical exponential correction, ensuring accurate estimation of both the crystal and magnetic structures. Examination of diffraction patterns between $1.5$~K, 15~K, 30~K, 45~K, and above the ordering temperature (55~K) reveals the presence of pure magnetic reflections at Q = $0.75$, $1.06$, $1.68$, $1.84$, $2.24$, $2.38$, $2.72$, $2.81$, $3.09$, $3.19$, $3.44$, $3.53$, $3.76$, $3.82$, $4.04$, $4.12$ and $4.62$~\AA$^{-1}$, as shown in FIG.~2(a), indicating a total of 17 magnetic peaks. No magnetic peak is found at high Q range as shown in FIG.~2(b). The refinement of the first four magnetic reflections is shown in FIG.~3(c). The refinements of the 5th to 17th magnetic reflections are presented in FIG.~3(d). To illustrate the detailed refinement of the magnetic peaks, the $5$th to $17$th peaks are presented in FIG.~4. FIG.~4(a) illustrates the temperature evolution of the $5$th to $17$th magnetic peaks, highlighting the change in their intensities at different temperatures. FIG.~4(b) displays all these magnetic peaks refinements in an enlarged view to show the goodness of fit.   The Miller indices corresponding to all $17$ magnetic peaks are listed in Supplementary TABLE S$1$ ~\cite{ref50}. These additional reflections violate the $ F$-centring extinction conditions and can be consistently indexed using the magnetic propagation vector $\mathbf{k} = (0,1,0)$~\cite{ref51}.

\begin{figure}[!ht]
    \centering
    \includegraphics[width=0.5\textwidth]{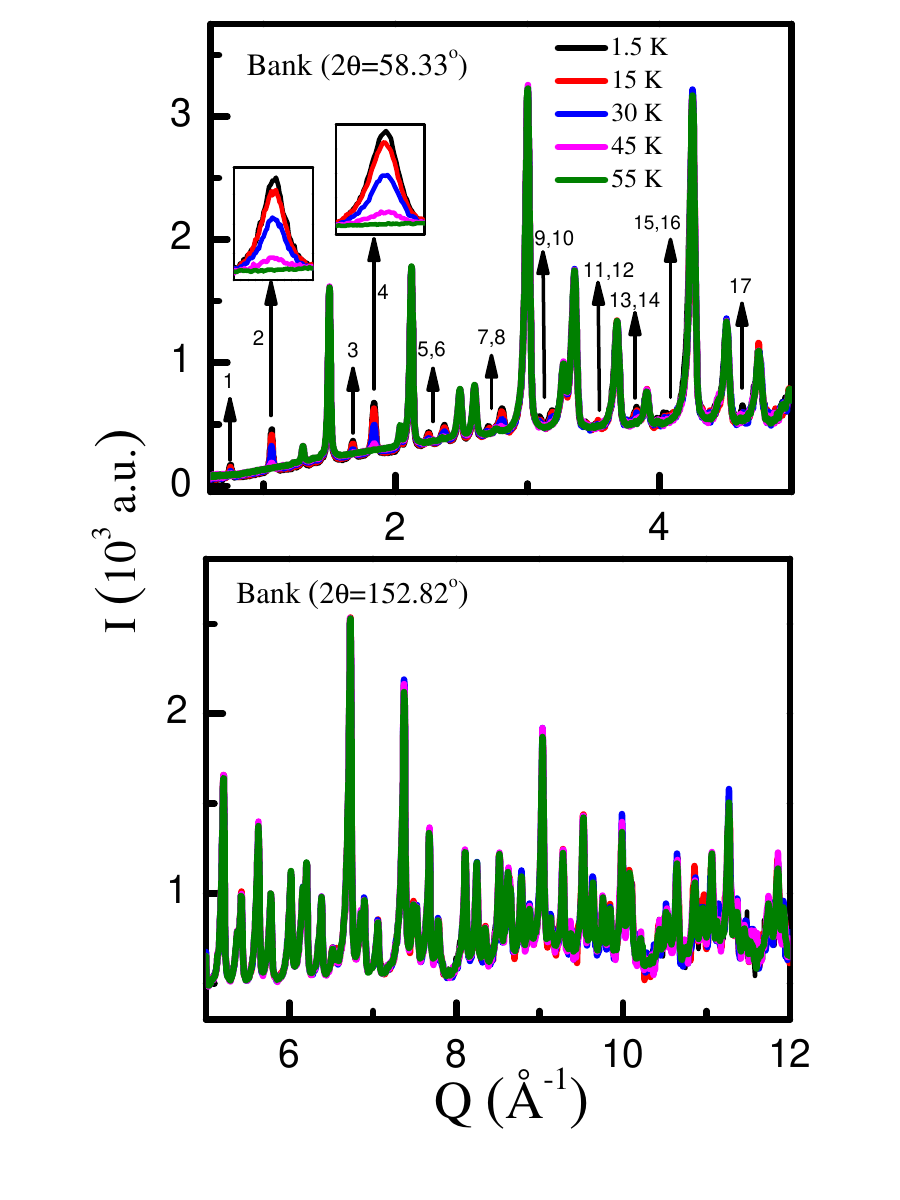} % Adjust filename and width
    \caption{ Comparison of neutron powder diffraction (NPD) patterns collected at 
1.5\,K, 15\,K, 30\,K, 45\,K, and 55\,K, with intensity $I$ plotted as a function of momentum transfer $Q$ (\AA$^{-1}$), highlighting the magnetic reflections from the WISH bank with average $2\theta = 58.33^{\circ}$. The inset shows an enlarged view of the two prominent magnetic peaks at $Q = 1.06$ and $1.84$~\AA$^{-1}$.
 }

    \label{Fig2}
\end{figure}

\begin{figure}[!ht]
    \centering
    \includegraphics[width=0.37\textwidth]{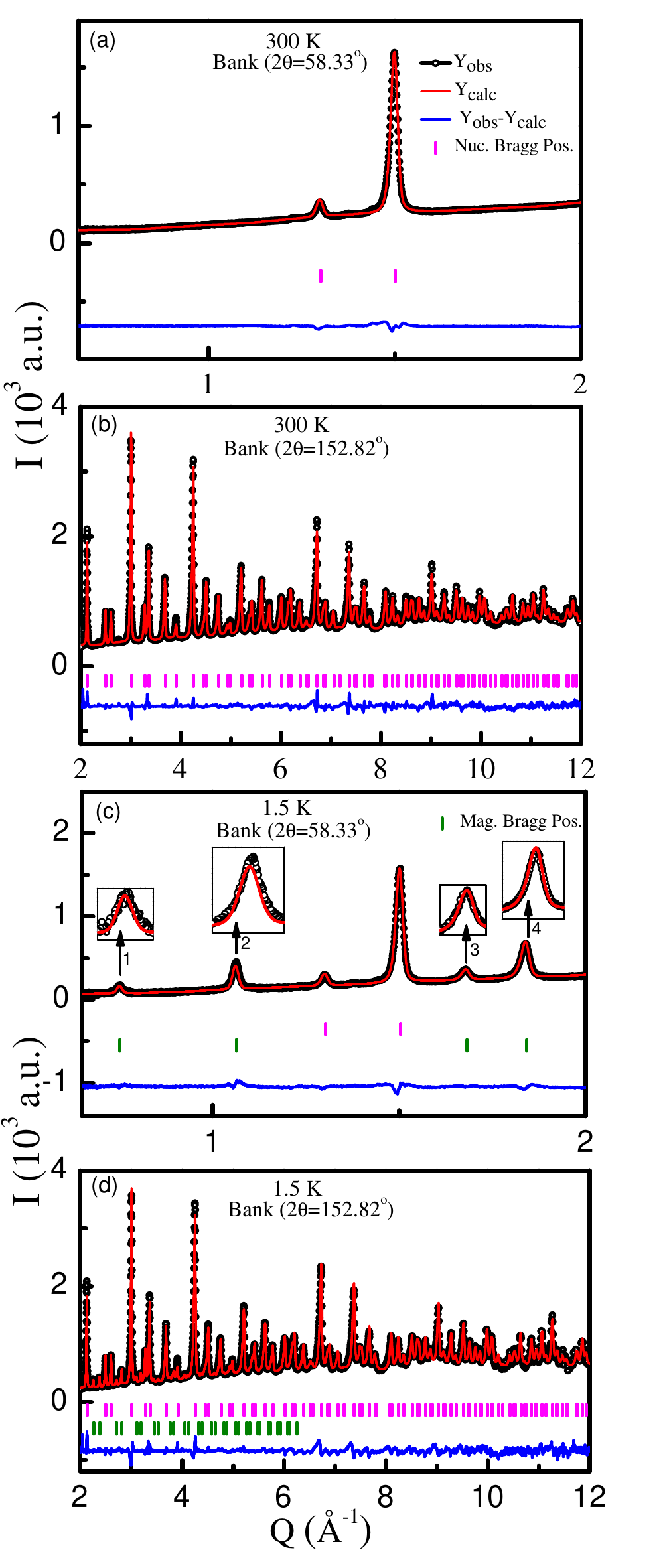} % Adjust filename and width
    \caption{Rietveld refinement of neutron diffraction data collected on the WISH diffractometer. Refinements at $300$~K in the range $0.65 \leq Q \leq 2~\mathrm{\AA^{-1}}$ using (a) Bank with average $2\theta = 58.33^{\circ}$ and (b) Bank with average $2\theta = 152.82^{\circ}$. Open black circles represent the observed data, the solid red line denotes the calculated profile, and the blue curve shows the difference between observed and calculated intensities. Vertical purple ticks mark the nuclear Bragg positions.(c) Rietveld refinement at $1.5$~K using the WISH Bank with average $2\theta = 58.33^{\circ}$. The first four magnetic Bragg peaks are observed at $Q = 0.75$, 1.06, 1.68, and $1.84~\mathrm{\AA^{-1}}$, as indicated in the figure. Vertical green ticks indicate magnetic Bragg positions.(d) Rietveld refinement at $1.5$~K using the WISH Bank with average $2\theta = 152.82^{\circ}$. The 5th ($Q = 2.24~\mathrm{\AA^{-1}}$) to 17th ($Q = 4.62~\mathrm{\AA^{-1}}$) magnetic Bragg peaks are observed in this range. Detailed fitting of individual peaks and the corresponding Miller indices ($hkl$) are provided in the Supplemental Material ~\cite{ref50}. 
}

    \label{Fig2}
\end{figure}

\begin{table}[h!]
    \renewcommand{\arraystretch}{0.8}
    \setlength{\tabcolsep}{4pt}
    \centering
    \caption{Atomic coordinates of Ba$_2$DyRuO$_6$ (space group Fm-$3$m) at 1.5 K and 300 K.}

    % ====== 1.5 K ======
    \vspace{2mm}
    \textbf{1.5 K} \\
    (Cell parameters: $a=b=c=8.3434(2)$\,\AA, $\alpha=\beta=\gamma=90^\circ$)

    \vspace{2mm}
    \begin{tabular}{|l|ccc|c|c|c}
        \hline\hline
        Atom (Site) & x & y & z & B$_{\text{iso}}$ & Occupancy \\
        \hline
        O ($24$e)  & 0.2349(3) & 0 & 0 & 0.7066(4) & 1 \\
        Ru($4$a) & 0 & 0 & 0 & 0.2241(4) & 0.1664(1) \\
        Ba ($8$c) & 0.25 & 0.25 & 0.25 & 0.3764(1) & 0.3328(2) \\
        Dy ($4$b) & 0.5 & 0.5 & 0.5 & 0.2241(2) & 0.1664(1) \\
        \hline
    \end{tabular}

    \vspace{4mm}

    % ====== 300 K ======
    \textbf{300 K} \\
    (Cell parameters: $a=b=c=8.3562(4)$\,\AA, $\alpha=\beta=\gamma=90^\circ$)

    \vspace{2mm}
    \begin{tabular}{|l|ccc|c|c|c}
        \hline\hline
        Atom (Site) & x & y & z & B$_{\text{iso}}$ & Occupancy \\
        \hline
        O ($24$e)   & 0.2351(3) & 0 & 0 & 1.1888(5) & 1 \\
        Ru($4$a) & 0 & 0 & 0 & 0.3275(3) & 0.1664(1) \\
        Ba ($8$c) & 0.25 & 0.25 & 0.25 & 0.9252(2) & 0.3328(2) \\
        Dy ($4$b) & 0.5 & 0.5 & 0.5 & 0.3275(3) & 0.1664(1) \\
        \hline
    \end{tabular}

\end{table}

\begin{figure}[!ht]
    \centering
    \includegraphics[width=0.52\textwidth]{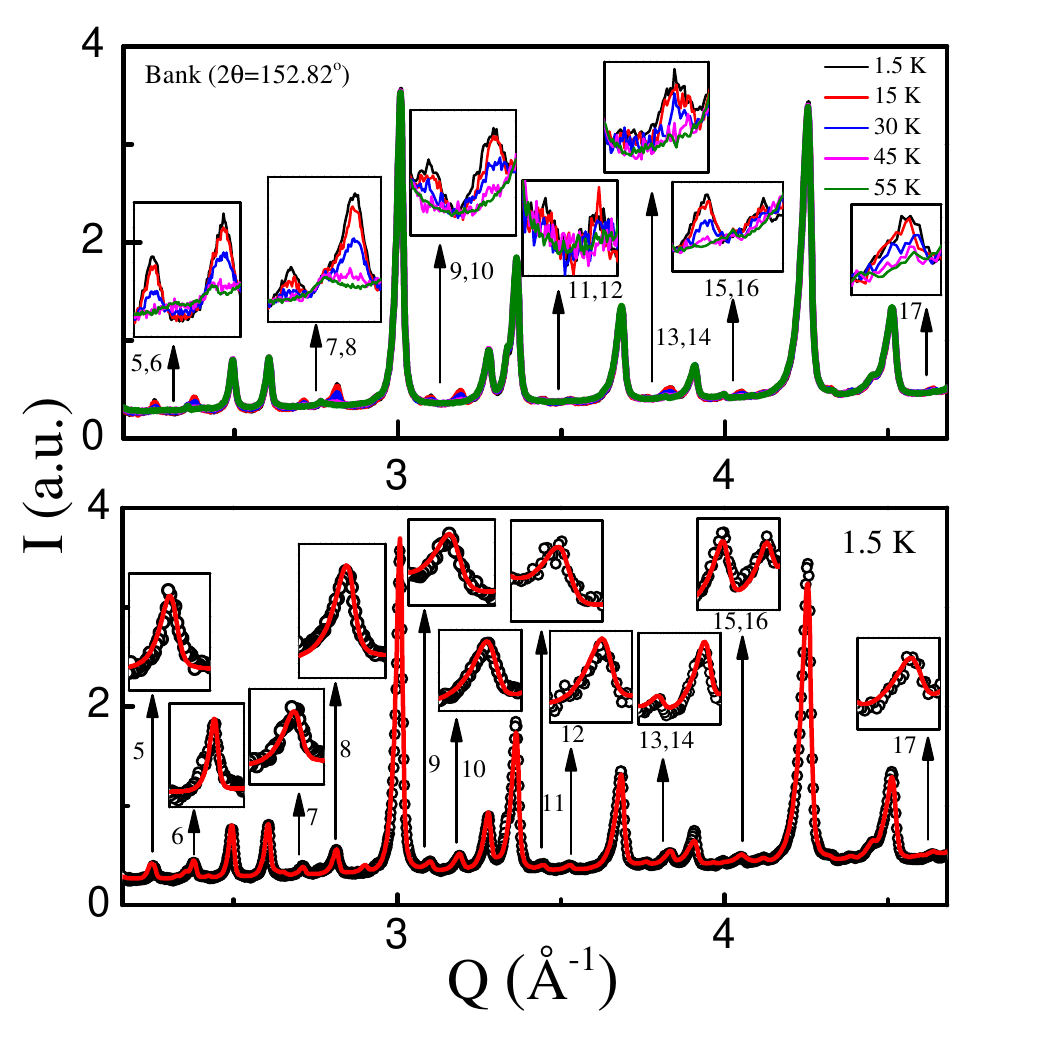} % Adjust filename and width
    \caption{(a) Comparison of neutron powder diffraction (NPD) patterns measured at $1.5$, $15$, $30$, $45$, and $55$ K for Bank ($2\theta = 152.82^\circ$), plotted as intensity $I$ versus momentum transfer $Q$ (\AA$^{-1}$), highlighting the 5th–17th magnetic peaks (top panel). Insets show the temperature evolution of individual magnetic peaks. (b) Enlarged view at $1.5$ K showing the Rietveld refinement of the 5th–17th magnetic peaks, demonstrating the quality of the fit.
}

    \label{Fig2}
\end{figure}

The cubic metric of the unit cell does not allow a unique determination of the moment direction with respect to the nuclear structure from powder data. The only reliable information that can be extracted is whether the magnetic moments lie parallel or perpendicular to the propagation vector. Preliminary refinements indicate that the moments lie in the plane perpendicular to the propagation vector $\mathbf{k}$. To further investigate the moment orientation, we examined the magnetic space groups derived from the parent space group Fm-$3$m (No.~225)~\cite{ref52}. Among the magnetic space groups obtained from a single irreducible representation and a single arm of the star of $\mathbf{k}$, only four groups—$P_I4/mnc$, $C_Amca$, $P_lnnm$ and $P_a2_1/c$—permit nonzero magnetic moments on both Ru$^{5+}$ and Dy$^{3+}$ sites. The $P_I4/mnc$ subgroup corresponding to the mX3$^+$ irreducible representation allows a moment parallel to the propagation vector (in our case along the $b$ direction of the parent cubic structure), $C_Amca$ allows a moment along the $a$ axis, and $P_lnnm$ allows a moment along the [110] direction within the $ac$ plane of the parent structure, whereas the $P_{a}2_1/c$ magnetic space group allows the moment to lie in a general direction within the $ac$ plane of the cubic structure. The latter three space groups correspond to the mX5$^+$ irreducible representation with different order-parameter directions. The magnetic refinement shows that the moments cannot be fitted exclusively along the $b$ axis, thus ruling out the $P_I4/mnc$ subgroup. Regarding the remaining three magnetic space groups, these cannot be distinguished from the powder data due to the cubic metric of the unit cell. For this reason, we have arbitrarily chosen to place the moment along the [110] direction of the parent cubic unit cell, corresponding to the $P_lnnm$ magnetic space group. The latter is defined in a unit cell related to the cubic parent structure by the transformation  \[
\{ (0,1,0),\, (-\tfrac{1}{2},0,\tfrac{1}{2}),\, (\tfrac{1}{2},0,\tfrac{1}{2}) \}
\] with the same origin choice. If the equivalent propagation vectors $\mathbf{k} = (1, 0, 0)$ or $\mathbf{k} = (0, 0, 1)$ are considered, equally good fits are obtained for moment orientations within the $bc$ and $ab$ planes, respectively, corresponding to the different domains of same magnetic space group $P_{l}nnm$. The magnetic structure is shown in FIG.~5 for  $\mathbf{k} = (0, 1, 0)$  using the parent structure unit cell for clarity. The magnetic moments at 1.5~K are found to be $\mu_{\text{Ru}^{5+}} = 1.6(1)\ \mu_B$ and $\mu_{\text{Dy}^{3+}} = 5.1(1)\ \mu_B$, obtained from the Rietveld refinement as shown in FIG.~3(c) and (d).

\begin{figure}[h!]
    \centering
    \includegraphics[width=0.65\textwidth]{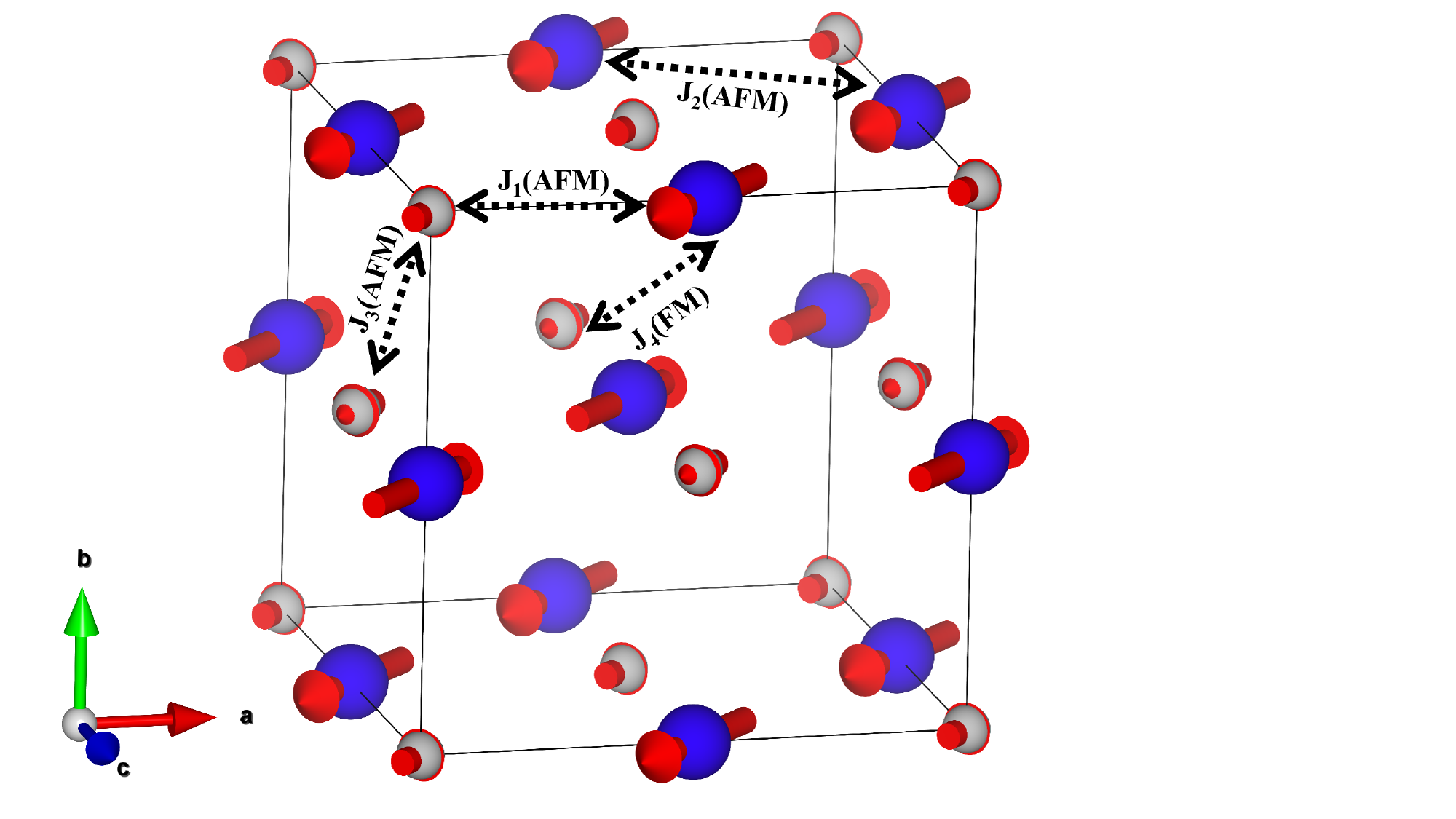} 
    \caption{Magnetic ground state of Ba$_2$DyRuO$_6$ at 1.5 K, drawn in the cubic unit cell of the parent structure showing a collinear magnetic structure, showing one symmetry-allowed within the $ac$ plane with propagation vector $\mathbf{k} = (0, 1, 0)$. Owing to the cubic metric and powder averaging, the absolute moment direction is not uniquely determined by the neutron powder diffraction data, and equivalent domain-related orientations give indistinguishable fits. Blue spheres represent Dy atoms, and Grey spheres represent Ru atoms. The dominant exchange path is indicated by black dotted lines.}
    \label{fig:fig3}
\end{figure}

\begin{figure}[h!]
    \centering
    \includegraphics[width=0.45\textwidth]{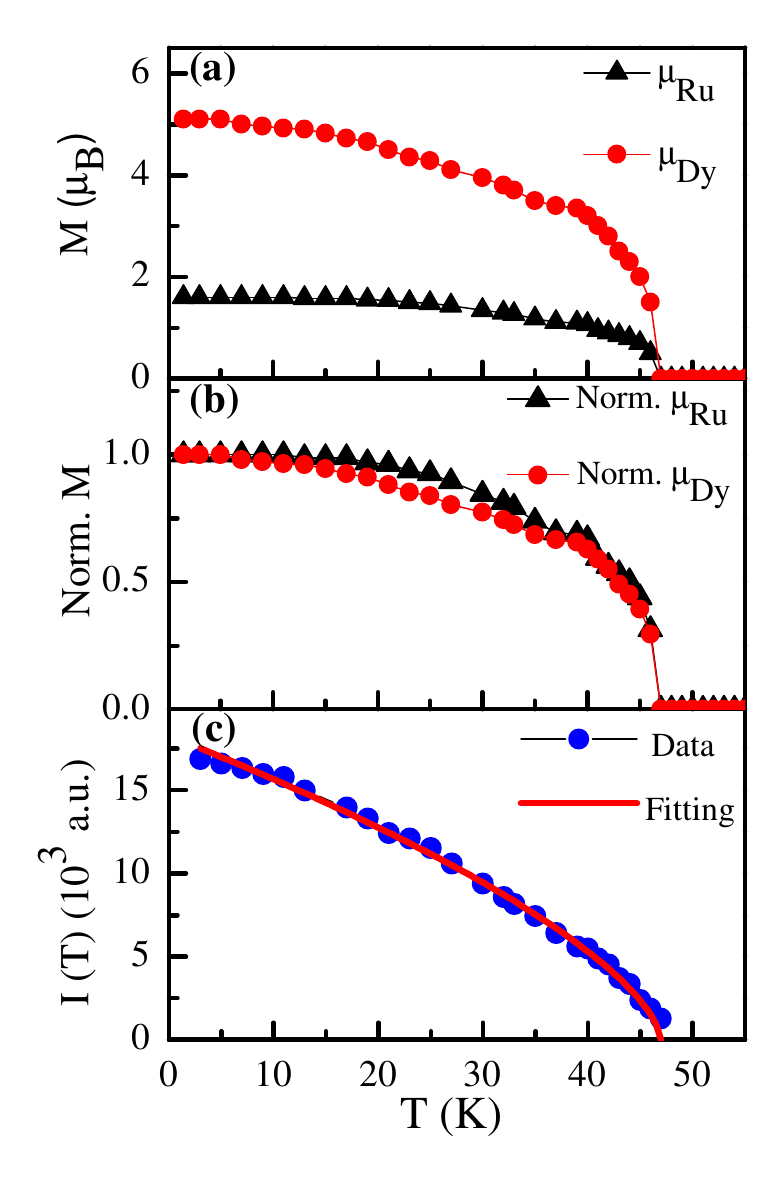}
    \caption{Thermal variation of (a) Dy$^{3+}$ and Ru$^{5+}$ moments based on refinement, 
    (b) normalised moments of Dy$^{3+}$ and Ru$^{5+}$, and 
    (c) temperature dependence of the integrated intensity \(I(T)\) below the ordering temperature. 
    A uniform uncertainty of $\pm 0.1\,\mu_{\mathrm{B}}$ is considered for the refined magnetic moments. 
    The data are fitted with \( I(T) = I_0(1 - T/T_N)^{2\beta} \) using \( \beta = 0.325(2) \) (Ising model).}
    \label{fig:fig3}
\end{figure}

   For the monoclinic ($P2_1/n$) compound Ba$_2$PrRuO$_6$, the Ru and Pr moments lie in the $ac$ plane, but the moments are non-collinear~\cite{ref31}. In Ba$_2$PrRu$_{0.9}$Ir$_{0.1}$O$_6$, the Pr moments align along the $c$ axis, whereas the Ru moments lie within the $ac$ plane~\cite{ref31}. In Ba$_2$HoRuO$_6$~\cite{ref30}, the magnetic moments are oriented along the $c$ direction. Even when only the $A$ site of the double perovskite $A_2BB'$O$_6$ is substituted, substantial changes are observed in both the crystal and magnetic structures. For example, in Sr$_2$DyRuO$_6$ (monoclinic space group $P2_1/n$)~\cite{ref34}, the moments align along the $b$ direction. In Sr$_2$HoRuO$_6$ (also $P2_1/n$)~\cite{ref33}, both Ru and Ho moments align along the $c$ direction. In Sr$_2$ErRuO$_6$, the magnetic structure features Er$^{3+}$ and Ru$^{5+}$ moments that remain predominantly aligned along the $c$ axis while exhibiting a slight tilt away from perfect collinearity~\cite{ref35}. Interestingly, in Sr$_2$YbRuO$_6$~\cite{ref37}, the moments of both Yb and Ru are arranged in a coplanar configuration within the $ab$ plane. These findings suggest that substituting either the $A$-site alkaline-earth ions or different rare-earth elements exerts a profound influence on the orientation and interaction of the Ru and rare-earth magnetic moments.

The temperature dependence of the magnetic moment variation of Ru$^{5+}$ and Dy$^{3+}$, obtained from Rietveld refinement of time-of-flight neutron diffraction data, is shown in FIG.~6(a), with the normalised moment in FIG.~6(b). These results confirm the magnetic ordering ($T_N$) $47(1)$ K, consistent with the value reported for bulk ~\cite{ref29}. In Ba$_2$DyRuO$_6$, the Ru$^{5+}$ moments saturate earlier than the Dy$^{3+}$ moments. (see FIG. $6(b)$) . This result is consistent with other similar Ru-based double perovskite systems~\cite{ref34}. FIG.~$6(c)$ shows the plot of the integrated intensity (I) of the strongest magnetic peak (peak located at $Q = 1.84 \,\text{\AA}^{-1}$ from FIG.~2(a)) at various temperatures below the ordering temperature. The estimated critical exponent $\beta = 0.325(2)$  is compatible with a 3D Ising-nature of moments~\cite{ref53}. This suggests a possible Ising-like character.

\subsection{Inelastic Neutron Scattering}

To understand phonon and magnon excitations, the crystal electric field (CEF), and to interpret the magnetic ground state, we performed inelastic neutron scattering (INS) experiments at various incident energies ~\cite{ref55}. FIG.$7$(a)--(c) color contour plots describing low-energy excitations in the $Q$ region $0.4$ to $1.2~\text{\AA}^{-1}$, recorded at temperatures of $5~\mathrm{K}$, $30~\mathrm{K}$, and $100~\mathrm{K}$ with an incident energy $E_i = 29.7~\mathrm{meV}$. The $1$D plot in FIG. $7$ (d) displays excitations at around $2.9$, $4.9$, and $9.6~\mathrm{meV}$, which diminish at $30~\mathrm{K}$ and completely vanish above the ordering temperature at $100~\mathrm{K}$, indicating the magnetic origin of these features. Further, the intensity of these low-energy excitations decreases with increasing Q, as the magnetic form factor decreases with increasing Q. Hence, these excitations are characterised as magnon excitations. Such a temperature dependence is characteristic of collective magnetic (magnon) excitations (see FIG. $7 $(d)), which vanish above the magnetic ordering temperature, whereas crystal electric field (CEF) excitations typically persist well above the ordering temperature ~\cite{ref54}. Further, the dispersive nature of these excitations is discussed and elaborated in FIGs.~S4 and S5 in the Supplementary Material \cite{ref50}.

\begin{figure}[h]
\centering
\includegraphics[width=.5\textwidth]{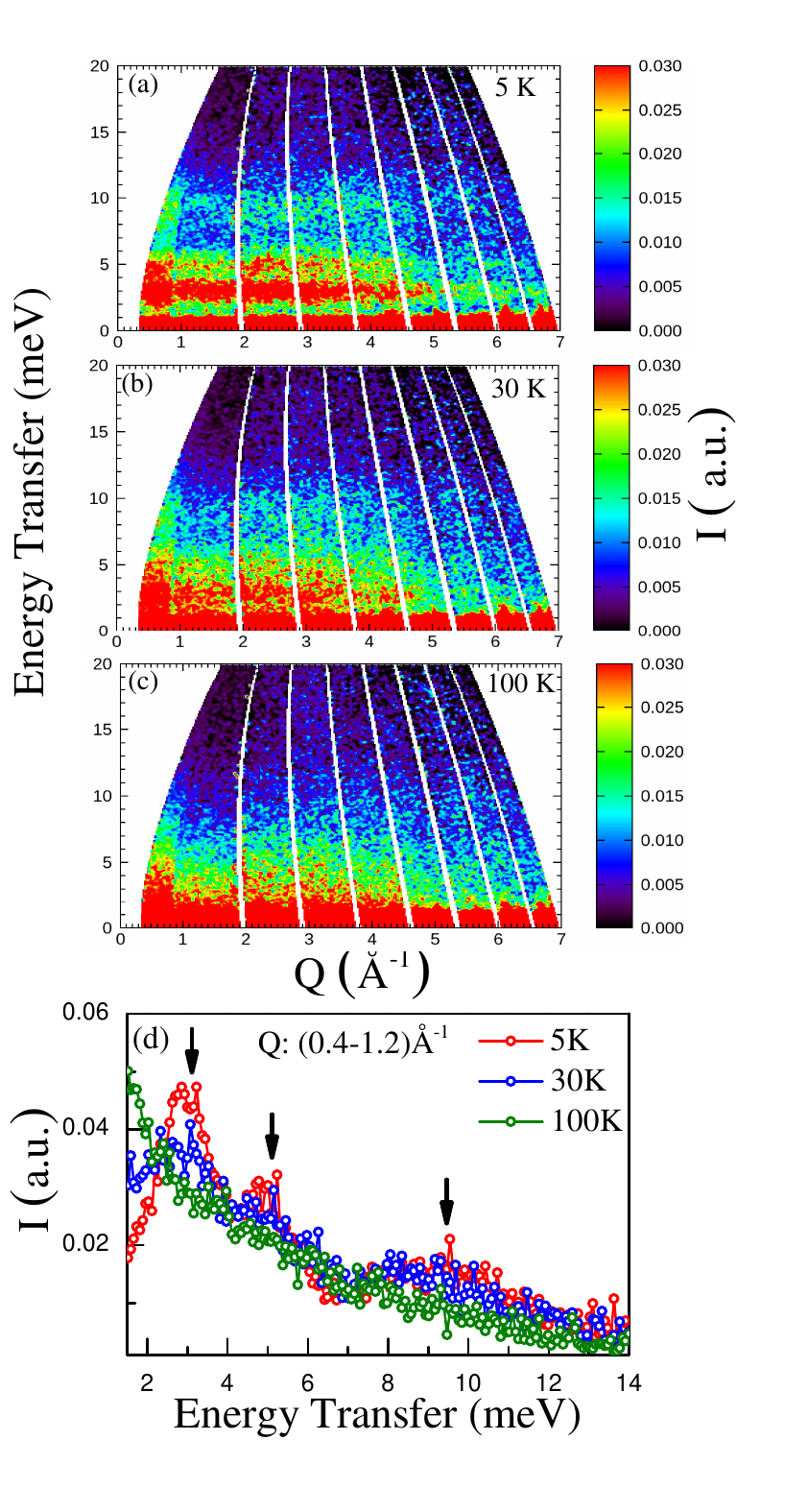} 
\caption{%
(a--c) Color contour  plots for $E_i = 29.7$ meV showing energy excitations at $5$ K, $30$ K, and $100$ K, respectively. (d) Comparison of magnon excitations (Intensity(I) vs.\ Energy Transfer) at various temperatures in the $Q$ region $0.4$ to $1.2~\text{\AA}^{-1}$.%
}
\label{fig:5}
\end{figure}

We next examine the crystal electric field (CEF) excitations arising from Dy$^{3+}$. The contour plots of INS spectra with an incident neutron energy of $E_i = 180~\mathrm{meV}$ are shown in FIG. $8$ (a)--(c) in the $Q$ region $1.5$ to $3~\text{\AA}^{-1}$ for temperatures of $5~\mathrm{K}$, $30~\mathrm{K}$, and $100~\mathrm{K}$, respectively. At $100~\mathrm{K}$, excitations are observed at $37.5~\mathrm{meV}$ and $71.8~\mathrm{meV}$. The $1$D plot in FIG.~8(d) shows the CEF transitions at 5, 30, and 100~K. At $5~\mathrm{K}$, the $37.5~\mathrm{meV}$ excitation shifts to $46.5~\mathrm{meV}$, possibly due to Zeeman splitting. Similar kind CEF excitations have also been reported for Dy$^{3+}$ in the Sr$_2$DyRuO$_6$ system~\cite{ref34}. The $Q$-dependence of the CEF excitation intensity follows a decay, as demonstrated in supplementary FIGs.~S1 and S2 ~\cite{ref50}, confirming the magnetic origin of these excitations associated with Dy$^{3+}$ \cite{ref56}.

FIG.~9 presents a one-dimensional plot showing phonon excitations observed in the high-$Q$ region, consistent with the phonon spectral behaviour where the phonon intensity increases with $ Q$ following $I \propto Q^{2}$. FIG.$9$(a) displays excitations around $9$--$28~\mathrm{meV}$, $32$--$38~\mathrm{meV}$, $43~\mathrm{meV}$, $51~\mathrm{meV}$, and $66$--$74~\mathrm{meV}$ for an incident energy of $E_i = 100~\mathrm{meV}$ at $5~\mathrm{K}$. For comparison and to probe higher energy ranges, FIG. $9$ (b) shows data with $E_i = 180~\mathrm{meV}$, revealing the same phonon excitations along with a weak feature near $90~\mathrm{meV}$.

\begin{figure}[h]
\centering
\includegraphics[width=0.5\textwidth]{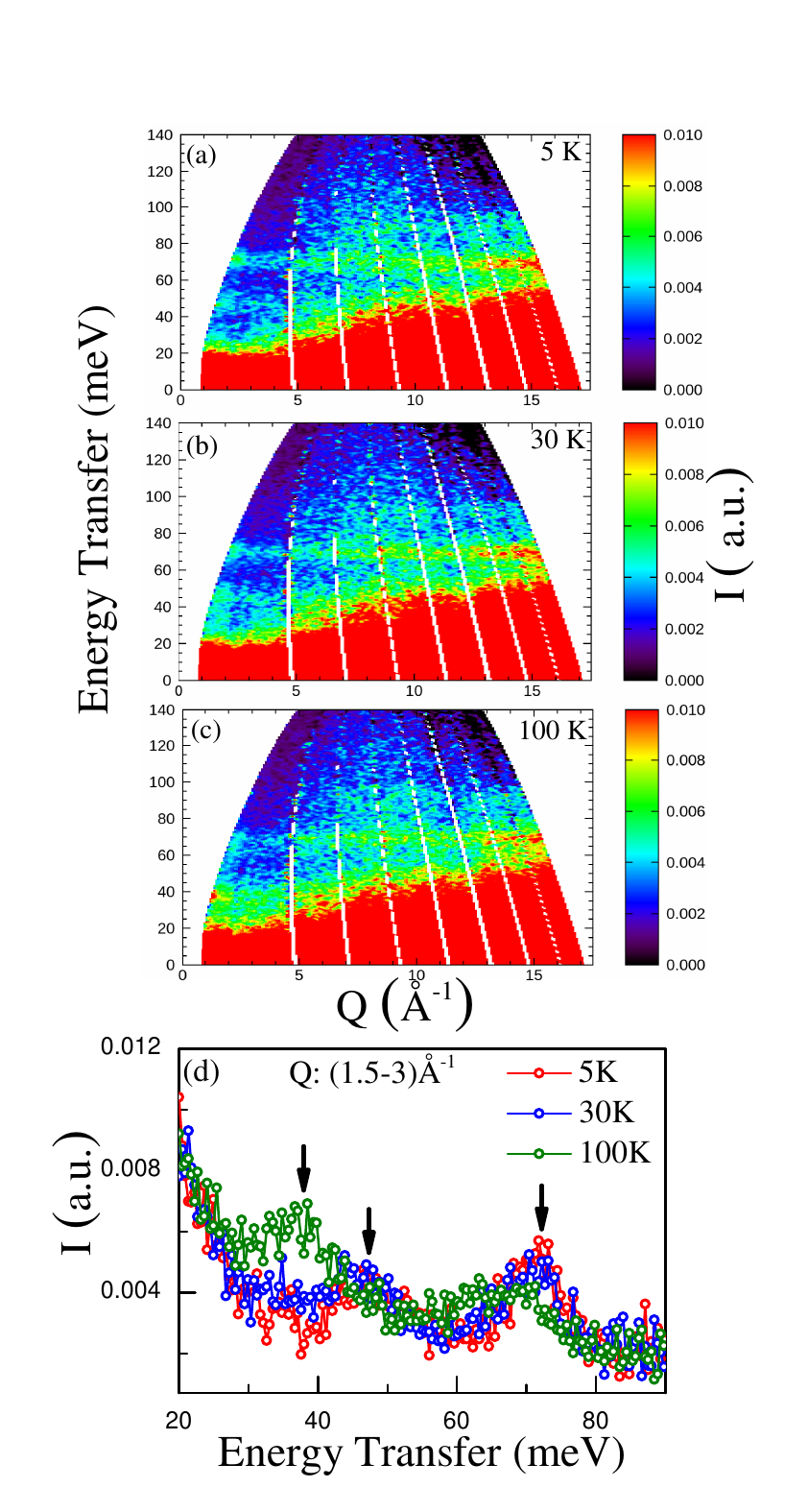} % Ensure file is in the same folder
\caption{%
 (a--c) Color contour plots for $E_i = 180$ meV showing excitation at 5~K, 30~K, and 100~K. (d) Comparison of CEF excitations (Intensity(I) vs.\ Energy Transfer) at various temperatures in the $Q$ region $1.5$ to $3.0~\text{\AA}^{-1}$.%
}
\label{fig:6}
\end{figure}

\begin{figure}[h]
\centering
\includegraphics[width=0.48\textwidth]{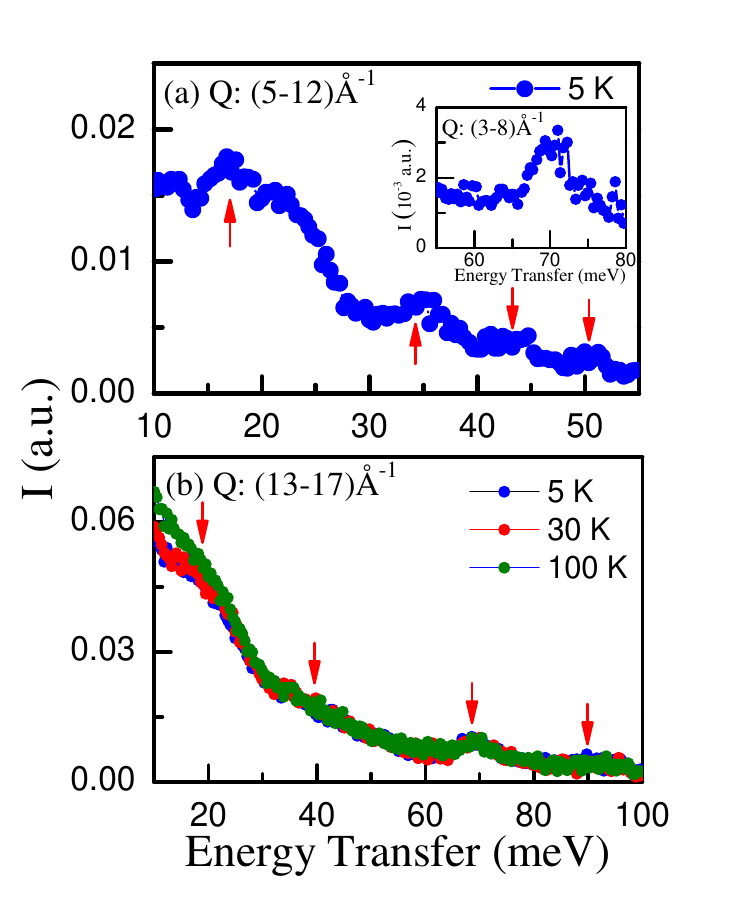} 
\caption{%
(a) Phonon excitations (Intensity(I) vs.\ Energy Transfer) measured with 
$E_i = 100~\text{meV}$ at $T = 5~\text{K}$ in the $Q$ range 
$5 \leq Q \leq 12~\text{\AA}^{-1}$. The inset shows phonon excitations 
in the energy range $55$--$80~\text{meV}$ for $3 \leq Q \leq 8~\text{\AA}^{-1}$.
(b) Comparison of phonon excitations (intensity vs.\ energy transfer) measured 
with $E_i = 180~\text{meV}$ at various temperatures in the$Q$ range 
$13 \leq Q \leq 17~\text{\AA}^{-1}$.
}
\label{fig:7}
\end{figure}

\subsubsection{SpinW: Theoretical modeling}
The spin–spin exchange interaction and the nature of anisotropy in the system are investigated by developing a theoretical model with \textsc{SpinW} to generate the magnon excitation spectrum, which is shown in FIG.$10$(a). 
The spin Hamiltonian of the Heisenberg system includes the exchange interaction $J$ and the single-ion anisotropy term $D$, expressed as
\begin{equation}
H = J \sum_{\langle i j \rangle} \vec{S}_i \cdot \vec{S}_j 
  + D \sum_{i} \left( S_i^{z} \right)^{2}.
\end{equation}
Here, $J$ represents the exchange interaction between neighbouring magnetic atoms, and $D$ denotes the anisotropy tensor~\cite{ref57}. From the simulations, the extracted $J$ and $D$ values are listed in TABLE II. The simulated magnon-dispersion curves and the corresponding contour plot of magnon spectra are shown in FIG.$10$(a) and FIG.$10$(b), respectively, which reveal clear excitations around $3$, $4$, and $8~\mathrm{meV}$ that closely agree with the experimentally obtained magnon excitations from INS (see FIG.$7$(a)). Due to the low intensity magnon feature from INS in Fig.$7$(a), it’s not possible here to exactly match the colour plot, although the energy levels are closely matched.

 \begin{figure}[h!]
    \centering
    \includegraphics[width=01\textwidth]{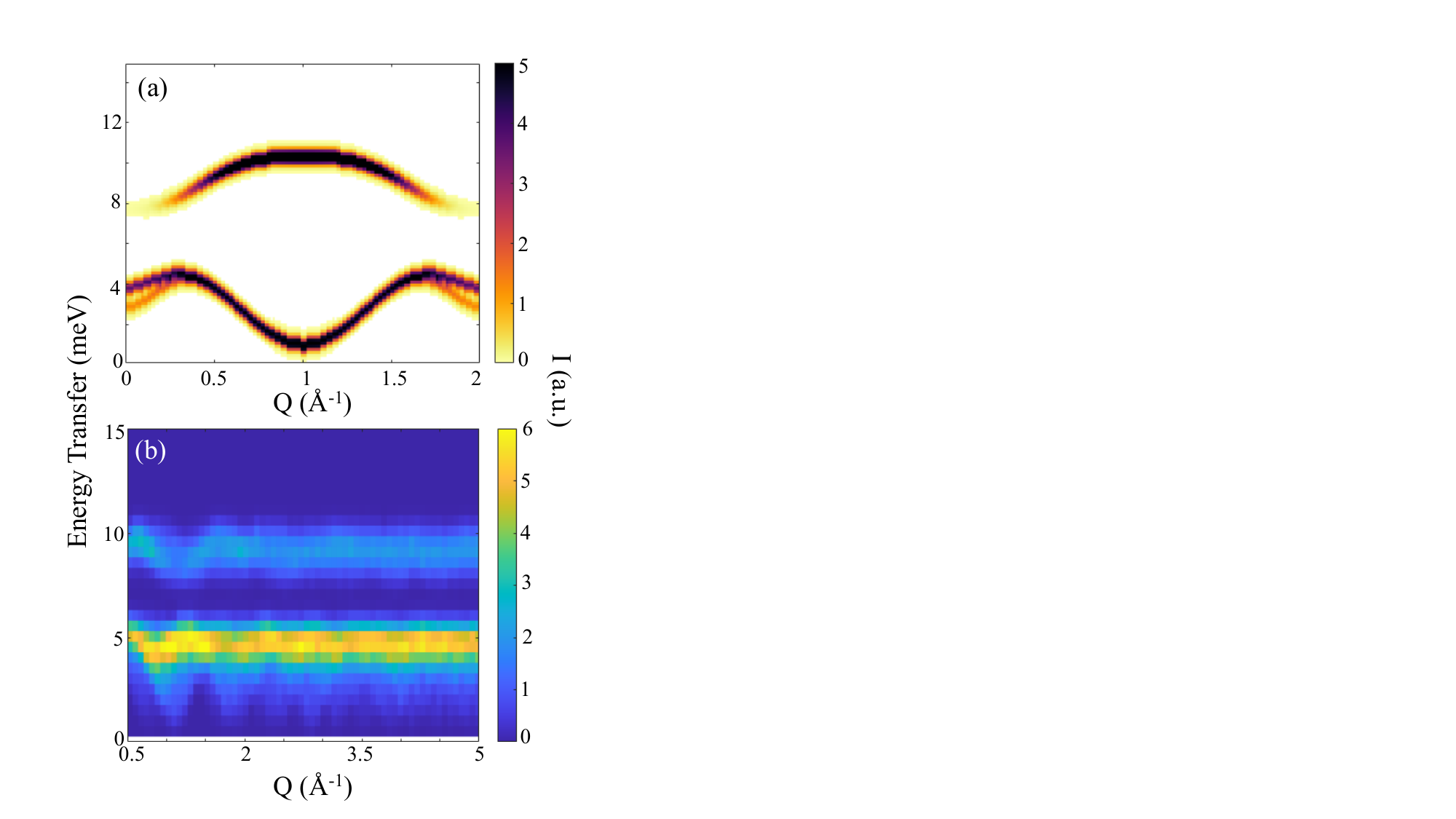}
    \caption{%
        (a) Spin-wave dispersion curve showing magnon excitations at approximately 3, 4, and 8~meV. 
        (b) Energy vs.\ momentum-transfer plot showing the 3D color contour of powder average spin-wave intensity obtained from \textsc{SpinW} simulations.%
    }
    \label{fig:fig7}
\end{figure}

 As shown in TABLE II, \( J_{\text{Dy--Ru}} > J_{\text{Ru--Ru}} > J_{\text{Dy--Dy}} \). In BDRO, the long-range magnetic order is dominated by the Ru--O--Dy antiferromagnetic (AFM) interaction, which is absent in similar compounds~\cite{ref34}, likely due to the large spatial separation and weak superexchange pathways. 
The Ru$^{5+}$ ions strongly hybridize with Dy$^{3+}$ through oxygen, forming a robust $180^\circ$ superexchange path. In contrast, the Ru--Ru interaction proceeds via a super--superexchange pathway (Ru--O--Dy--O--Ru), which is comparatively weaker. Similarly, the Dy--Dy interaction follows the Dy--O--Ru--O--Dy super--superexchange pathway, which is extremely weak because the Dy 4$f$ orbitals are highly localized, resulting in minimal overlap with the O 2$p$ states.

The anisotropy \(3 \times 3\) tensor contains only non-diagonal terms, specifically \(D_{zx}\) and \(D_{xz}\) (with \( \mathbf{k} = (0\,1\,0) \) and collinear Ising like moments lying in the \(ac\)-plane). Compared to the exchange interaction \(J\), the \(D\) terms are not strong enough to dominate the spin interactions. The anisotropy matrix contains nonzero off-diagonal elements (\(D_{xz} = D_{zx} = 0.10\)), which causes the Ising-like moments to preferentially align within the \(ac\)-plane. In this case, determining the exact moment direction is not possible: if the spins were constrained to the \(ab\) plane, the corresponding nonzero elements would be \(D_{xy} = D_{yx} = 0.10\); similarly, for moments lying in the \(bc\) plane, one would obtain \(D_{yz} = D_{zy} = 0.10\). Inclusion of the anisotropy term is therefore essential to reproduce the correct spin orientation and the observed magnon excitations. The exact reproducibility of INS spectra with this model, with an anisotropy tensor, further supports the Ising character of the spin system as predicted through neutron diffraction results.

\begin{table}[h!]
\centering
\caption{Exchange interactions and single-ion anisotropy used in the \textsc{SpinW} simulation, where nn and nnn refer to nearest and next-nearest-neighbour exchange, respectively. (Positive $J$ denotes antiferromagnetic interaction, while negative $J$ denotes ferromagnetic interaction.)}
\vspace{0.5em}
\begin{tabular}{|l|c|}
\hline\hline
\textbf{Parameter} & \textbf{Value (meV)} \\
\hline
\( J_1 \) (nn Dy–Ru)      & 0.52 \\
\( J_2 \) (nn Dy–Dy)       & 0.09 \\
\( J_3 \) (nn Ru–Ru)       & 0.12 \\
\( J_4 \) (nnn Dy–Ru)     & $-$0.10 \\

\hline
\end{tabular}

\vspace{1em}

\textbf{Anisotropy matrix \( D \) (meV):}
\[
D = \begin{bmatrix}
0 & 0 & 0.10 \\
0 & 0 & 0 \\
0.10 & 0 & 0
\end{bmatrix}
\]
\label{tab:J_D_matrix}
\end{table}

\subsubsection{CEF Point Charge Calculation of Dy$^{3+}$}

To interpret the crystal electric field (CEF) excitations theoretically, we employ point charge calculations. Here, we discuss the CEF splitting of Dy$^{3+}$ (4f$^9$), which is in the $O_h$ point group of $m\bar{3}m$ symmetry in cubic BDRO. For Dy$^{3+}$ (4f$^9$) in $O_h$ site symmetry, two doublets and four quartets are expected, yielding a total of 16 states (as $2J + 1 = 15/2 \times 2 + 1 = 16$)~\cite{ref58}. We now represent CEF excitation through the point charge model, where the CEF Hamiltonian for the Dy$^{3+}$ site symmetry is given below,

\begin{equation}
\hat{H}_{\mathrm{CEF}}
= B_{4}^{0}\,\hat{O}_{4}^{0}
+ B_{4}^{4}\,\hat{O}_{4}^{4}
+ B_{6}^{0}\,\hat{O}_{6}^{0}
+ B_{6}^{4}\,\hat{O}_{6}^{4}.
\end{equation}

The single-ion crystal electric field (CEF) interaction is parametrized as a linear combination of the four symmetry-allowed Stevens operators, \( \hat{O}_n^m \), for the \( O_h \) site symmetry of the DyO$_6$ octahedral environment~\cite{ref59}.

\begin{figure}[h!]
    \centering
    \includegraphics[width=0.6\textwidth]{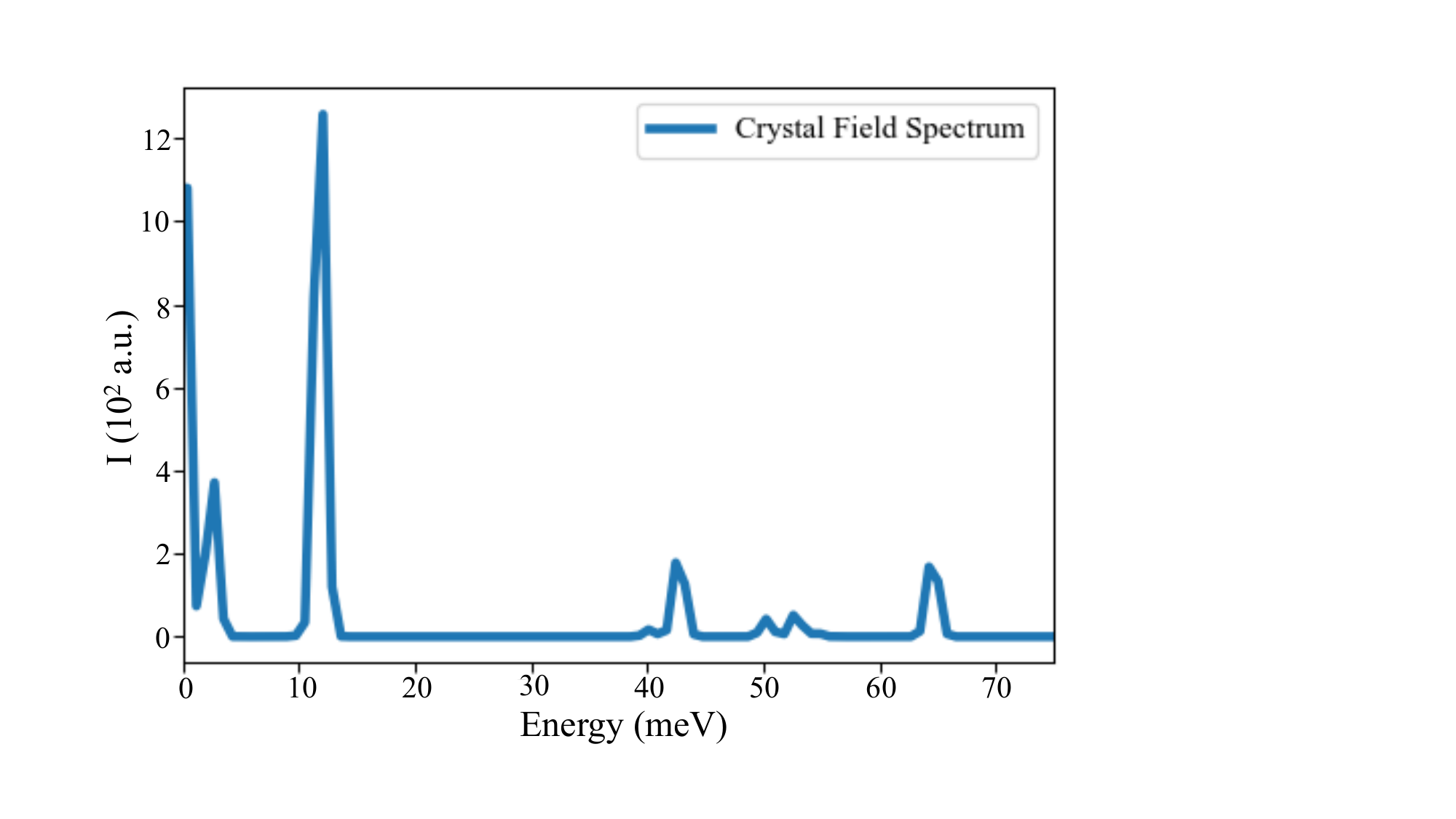}
    \caption{%
       Crystal field excitations calculated using the point charge model in the \textsc{Mantid} Crystal Field program at 100\,K, within the paramagnetic region, with intensity (I) plotted as a function of energy (E).
    }
    \label{fig:fig8}
\end{figure}
\begin{figure}[h!]
    \centering
    \includegraphics[width=0.7\textwidth]{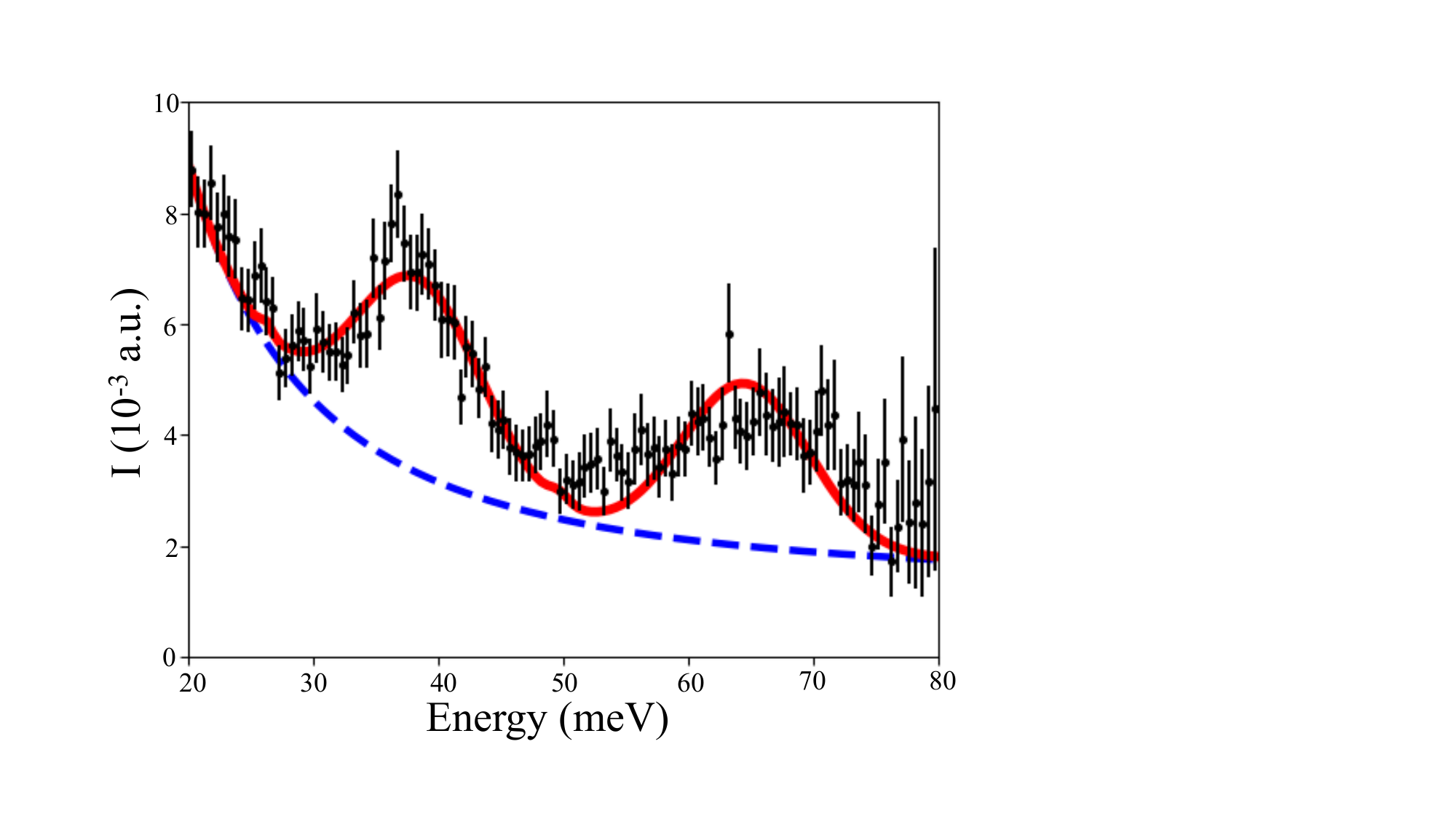}
    \caption{Crystal electric field (CEF) fit to the 5~K inelastic neutron scattering spectrum of Dy$^{3+}$ . The solid blue line represents the total fit consisting of the elastic peak and a linear background.}
    \label{fig:fig9}
\end{figure}

The calculated Stevens parameters, $B_4^0$ and $B_6^0$, obtained from the simulated results for Dy$^{3+}$ in Ba$_2$DyRuO$_6$, which are used to reproduce the measured CEF levels at 100~K, are listed in TABLE~III. For cubic symmetry, $B_4^4 = 5B_4^0$ and $B_6^4 = -21B_6^0$ are related~\cite{ref60}.

\begin{table}[h!]
\centering
\caption{Stevens parameters $B_n^m$ for Dy$^{3+}$ in Ba$_2$DyRuO$_6$ at 100~K and 5~K.}
\begin{tabular}{|c |c| c|}
\hline\hline
\textbf{Parameter} & \textbf{100 K (meV)} & \textbf{5 K (meV)} \\
\hline
$B_4^0$ & $-8.84(5)\times10^{-4}$ & $-1.43(6)\times10^{-3}$ \\
$B_6^0$ & $8.30(6)\times10^{-6}$  & $-1.22(5)\times10^{-5}$ \\
\hline\hline
\end{tabular}
\label{tab:stevens}
\end{table}

 The CEF excitations obtained from point-charge calculations reveal a ground-state doublet $\ket{0}$ at 0~meV, a first excited quartet $\ket{1}$ at 11.8~meV, a second excited doublet $\ket{2}$ at 14.3~meV, a third excited quartet $\ket{3}$ at 54.5~meV, and a fourth excited quartet $\ket{4}$ at 64.5~meV. The intermediate-state transitions at 100~K are also visible due to thermal population, as shown in FIG. ~11. The observed intermediate transitions correspond to the following states: $\ket{1}$  $\rightarrow$ $\ket{2}$ at 2.5~meV ((14.3--11.8)~meV), 
$\ket{1} \rightarrow \ket{3}$ at 42.7~meV ((54.5--11.8)~meV), 
$\ket{1} \rightarrow \ket{4}$ at 52.7~meV ((64.5--11.8)~meV), 
$\ket{2} \rightarrow \ket{3}$ at 40.2~meV ((54.5--14.3)~meV), 
and $\ket{2} \rightarrow \ket{4}$ at 50.2~meV ((64.5--14.3)~meV), as shown in FIG.~11. Among these, the 42.7~meV ($\ket{1} \rightarrow \ket{3}$) and 64.5~meV ($\ket{0} \rightarrow \ket{4}$) transitions closely match the experimentally observed CEF excitations at 46.5~meV and 71.8~meV, respectively (see FIG.~8(c)).

\begin{figure}[h!]
    \centering
    \includegraphics[width=1.2\textwidth]{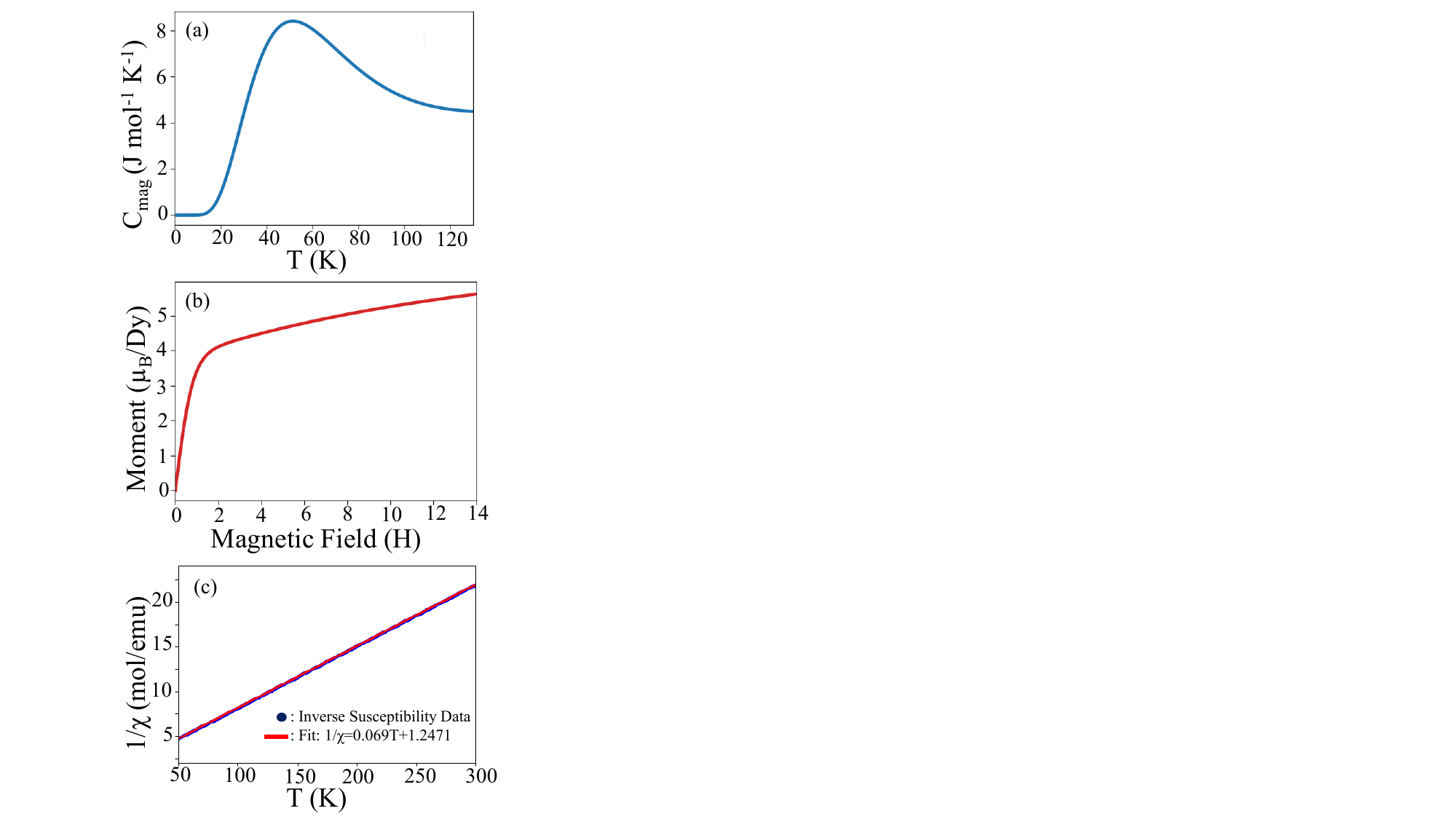}
    \caption{%
       Results from point charge theoretical calculations — (a) Magnetic heat capacity as a function of temperature $T$, (b) Dy$^{3+}$ magnetic moment calculated as a function of applied magnetic field $H$ , and (c) Curie–Weiss fit to the inverse magnetic susceptibility.
    }
    \label{fig:fig9}
\end{figure}

It is noted that the ordered state moment from the point charge model is calculated as \( \mu_{\text{ord}} = 5.7 \mu_B \) for Dy$^{3+}$, which is close to the experimental value of 5.1(1) $\mu_B$ (see FIG.~13(a)). The magnetic contribution of heat capacity, \( C_{\mathrm{mag}} \), obtained from this modelling is presented in FIG.~13(b), which shows a close agreement with the experimental \( C_{\mathrm{mag}} \) versus \( T \) data. Additionally, the Curie-Weiss fit of inverse susceptibility versus temperature, based on this theoretical calculation, yields \( \Theta_{\text{cp}} \approx -17.33 \) K and a paramagnetic moment \( \mu_{\text{Dy}} \approx 10.75 \mu_B \),(FIG.~13c) both of which are in close agreement with experimental results~\cite{ref29}. The ground state wavefunctions are found as,  
\[
|\psi_0\rangle = 0.23 \left| \mp\frac{11}{2} \right\rangle 
+ 0.45 \left| \mp\frac{3}{2} \right\rangle 
- 0.58 \left| \pm\frac{5}{2} \right\rangle 
- 0.63 \left| \pm\frac{13}{2} \right\rangle.
\] 

The point charge model is an approximation and does not provide an accurate ground eigenstate, but it depicts that the \(\left| \frac{13}{2} \right\rangle\) wavefunction contributes the most. It also predicts the energy gap of the crystal electric field (CEF) excitation between the Kramers doublets. In the paramagnetic region, we observe 8 CEF excitations as expected through point charge model as shown in FIG~10. The observation of only two CEF excitations experimentally may be attributed to the weak population of most CEF levels in the paramagnetic region and to thermal broadening, which can merge closely spaced excitations into broad peaks, thereby reducing their visibility in INS~\cite{ref61}.

Crystal electric field (CEF) fitting was carried out on the 5~K CEF excitations (see FIG.~8(a)) of the Dy$^{3+}$ ion in the DAaaS system, as shown in FIG.~12. The fitted Stevens parameters for $5$K are shown in Table III. Equivalently, within the Lea--Leask--Wolf (LLW) formalism, the corresponding parameters are
$x = 0.340 \pm 0.02$ and
$W = -0.257 \pm 0.01$~meV.

From the CEF fitting, a total of 16 energy levels are obtained, consistent with the $J = 15/2$ multiplet of Dy$^{3+}$. The level scheme consists of a quartet ground state, followed by a first excited doublet at 38.2~meV, a second excited quartet at 64.4~meV, a third excited quartet at 87.8~meV, and a fourth excited doublet at 111.8~meV. The excitations at 38.2 and 64.4~meV closely match the experimentally observed CEF excitations, as shown in Fig.~8(a) and Fig.~8(d). However, the $5$ K CEF excitations are measured below the magnetic ordering temperature, where the internal exchange field can lift or mix low-lying Kramers doublets; consequently, the ground state obtained from the $5$ K CEF fit can be interpreted as a quasi-quartet ground state formed by two closely spaced doublets rather than a true symmetry-protected quartet. The ground-state quartet wavefunctions, expressed in the $\lvert m_J \rangle$ basis, are given by

\[
|\psi_{0}\rangle =
-0.74 \left| \mp \dfrac{15}{2} \right\rangle
+ 0.58 \left| \mp \dfrac{7}{2} \right\rangle
+ 0.32 \left| \pm \dfrac{1}{2} \right\rangle
+ 0.03 \left| \pm \dfrac{9}{2} \right\rangle,
\]

\vspace{1em}

\[
\hspace{0.6cm}
=\pm 0.12 \left| \mp \dfrac{11}{2} \right\rangle
\mp 0.63 \left| \mp \dfrac{3}{2} \right\rangle
\mp 0.71 \left| \pm \dfrac{5}{2} \right\rangle
\pm 0.25 \left| \pm \dfrac{13}{2} \right\rangle .
\]

Although the ground state is predominantly composed of the $\lvert m_J \rvert = 15/2$ components, a significant admixture from lower $\lvert m_J \rvert$ states is observed. This indicates a deviation from the ideal Ising limit while retaining strong single-ion anisotropy for the Dy$^{3+}$ ion.

\subsection{Raman Spectroscopy}
In Ba$_2$DyRuO$_6$, the Wyckoff positions of Ba, Dy, Ru, and O are 8$c$, 4$b$, 4$a$, and 24$e$, respectively, corresponding to site symmetries $T_d$, $O_h$, and $C_{4v}$. Since the system possesses a centrosymmetric space group, the Raman-active modes are infrared-inactive.  The Raman active modes are summarised in Table IV, with the help of the Bibao crystallographic server~\cite{ref62}.

\begin{table}[h!]
\centering
\caption{Raman Active Modes}
\[
\Gamma = A_{1g} + E_{g} + 2T_{2g}
\]

\renewcommand{\arraystretch}{1.8} % adjust row height
\setlength{\tabcolsep}{14pt}      % column spacing

\begin{tabular}{@{}|l|c|c|c|@{}}
\hline\hline
\textbf{WP} & \textbf{$A_{1g}$} & \textbf{$E_{g}$} & \textbf{$T_{2g}$} \\
\hline
Ba (8c)  & $\cdot$ & $\cdot$ & 1 \\
Dy (4b)  & $\cdot$ & $\cdot$ & $\cdot$ \\
Ru (4a) & $\cdot$ & $\cdot$ & $\cdot$ \\
O   (24e) & 1       & 1       & 2 \\
\hline\hline
\end{tabular}
\end{table}

Here the $T_{2u}$ and $T_{1u}$ modes are inactive in Raman spectra but can be observed via hyper-Raman scattering. In FIG~13, our Raman spectra at 300~K show peaks at 341~cm$^{-1}$ (~42~meV), 443~cm$^{-1}$ (~55~meV), 555~cm$^{-1}$ (~69~meV), and 747~cm$^{-1}$ (~92~meV).  All these existing peaks are also observed in the INS data at 100 K. Based on comparison with the INS data, the peaks at 341~cm$^{-1}$ ($\sim$42~meV) and 555~cm$^{-1}$ ($\sim$69~meV) are assigned to crystal electric field (CEF) excitations (see FIG.~8(d)), with phonon modes also observable in this range. The peaks at 443~cm$^{-1}$ ($\sim$55~meV) and 747~cm$^{-1}$ ($\sim$92~meV) are attributed to pure phonon modes (see FIG.~9). 

\begin{figure}[h!]
    \centering
    \includegraphics[width=0.5\textwidth]{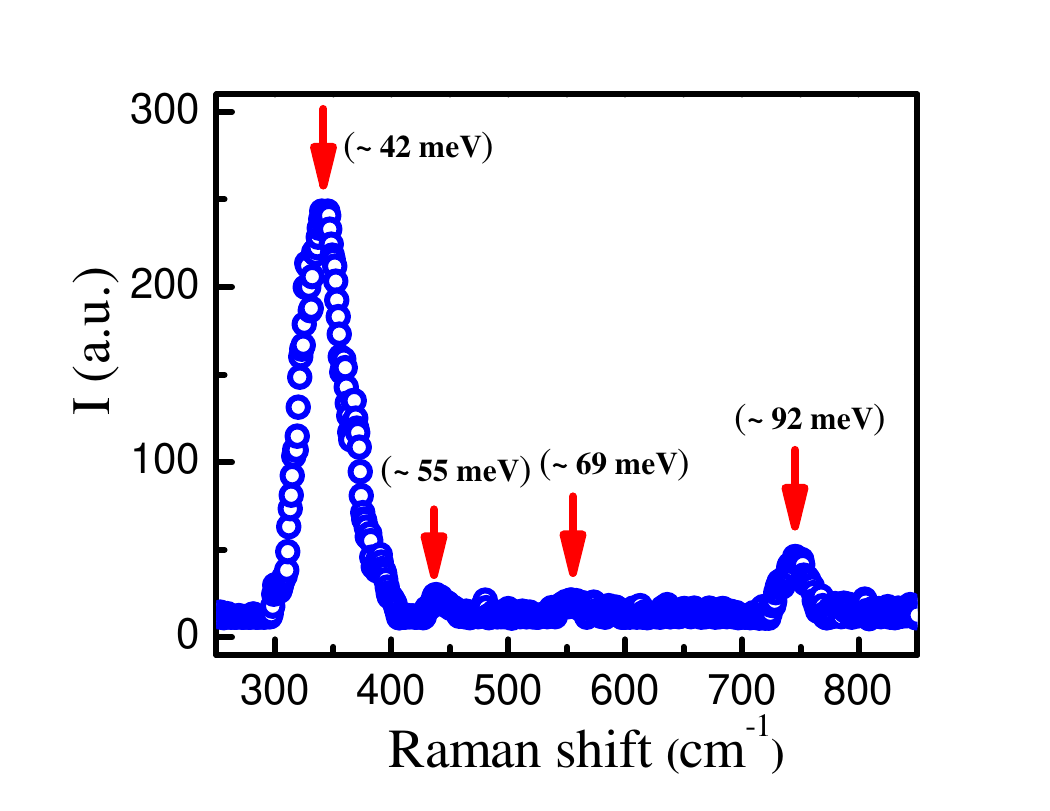}
    \caption{%
     Raman spectra of Ba$_2$DyRuO$_6$ recorded at 300\,K, with Intensity (I) plotted as a function of Raman shift (cm$^{-1}$), showing the characteristic vibrational modes.
    }
    \label{fig:fig10}
\end{figure}

\subsection{Machine Learning Models for Phonon Excitation}
We calculated the phonons across the full $|$Q$|$-range to separate the phonons with the magnetic signals seen in the INS data~\cite{ref47,ref48}. For this, we used the 5 K crystal structure and optimised the structure. FIG. 15(a) shows the calculated phonon spectrum that spans up to 25 THz and shows no imaginary frequencies, confirming the dynamic stability of the optimised structure. After getting this, we calculated the powder-averaged neutron scattering intensity convolved with the instrumental energy and momentum resolution. This calculation also included multiple phonon processes and temperature-dependent thermal population of phonons. 
\begin{figure}[!ht]
    \centering
    \includegraphics[width=0.9\textwidth]{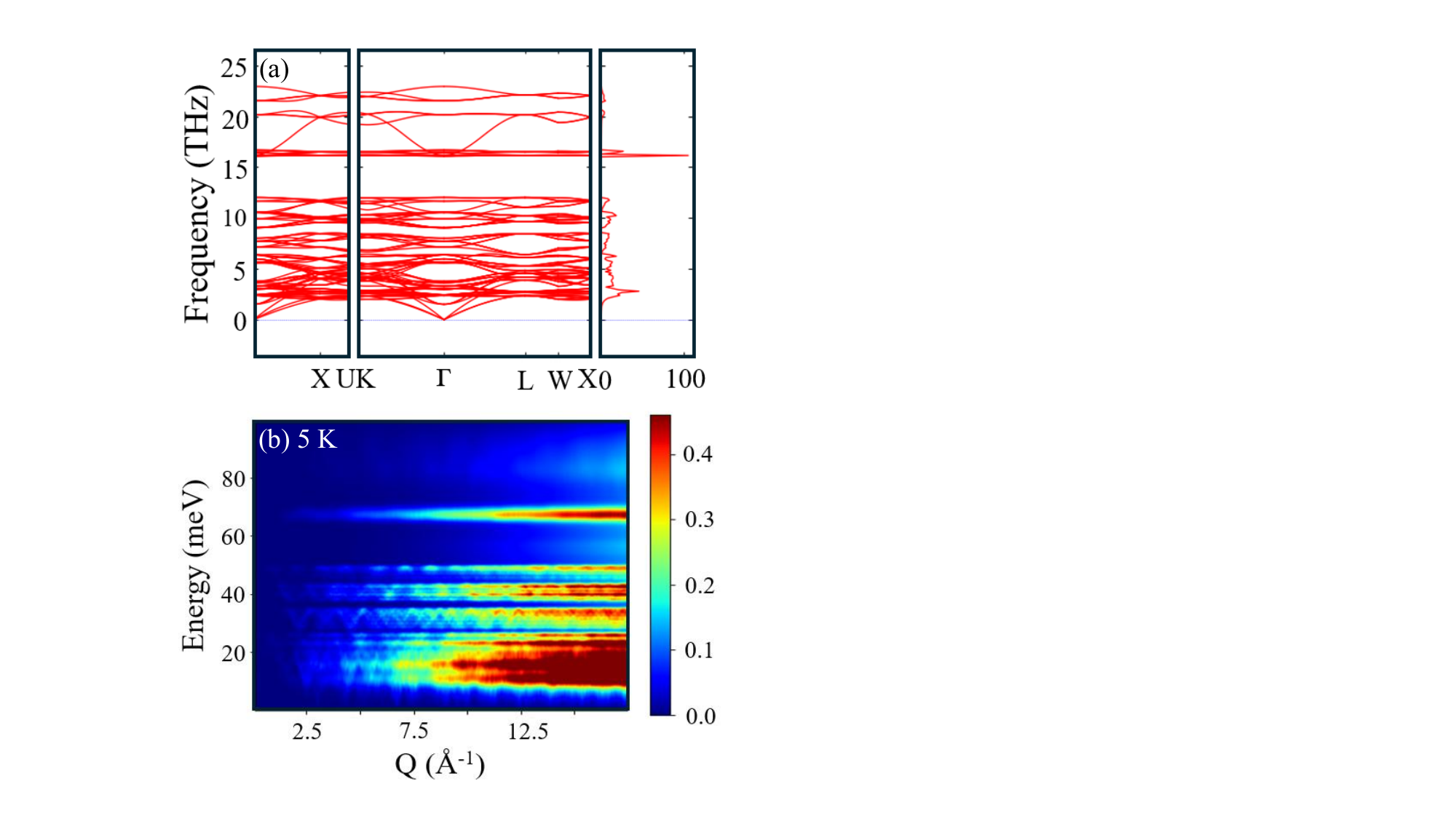}
    \caption{%
      (a) Calculated phonon modes for Ba$_2$DyRuO$_{6}$ based upon pre-trained machine learning force fields. Vertical axis is the phonon energy in units of THz. Panels indicate wave-vector dependence in reciprocal space using standard reciprocal space, (b) Calculated phonon scattering intensity at 5 K
    }
    \label{fig:fig14}
\end{figure}
The simulated intensity (FIG.$15(b)$) shows that phonon scattering dominates the $|$Q$|$ region $7.4$ to $17~\text{\AA}^{-1}$, consistent with the measured phonon cross sections. This calculation shows the low-energy excitations that appear broad and unresolved in the experimental INS data below $52 meV$ (see FIG. 9(a)), due to both the closely spaced modes and the instrumental energy resolution in this vicinity of energy transfer. The simulated spectrum allows us to identify well-separated phonon excitations in the range $9-28$ meV, $32-38$ meV, $43$ meV, $51$ meV, and $66-74$ meV, along with an additional weaker excitation near 90 meV. These calculated energies match very well with the excitations observed in the experimental spectra (see FIG. $9$), confirming both the accuracy of the phonon calculation and its consistency with the experimental INS results. Such a clear agreement between the INS experimental phonon spectra and the theoretical MLFF simulated phonon spectra is also observed in another ruthenate system~\cite{ref57}. ML-based force fields provide an efficient route to dense sampling of the phonon spectrum and enable direct comparison with INS intensity over broad Q–E ranges, without implying that such analyses would be unattainable using conventional approaches.

\section{Conclusion}
Experimental and theoretical investigations reveal the magnetic ground state of Ba$_2$DyRuO$_6$ and the mechanisms underlying its magnetic and lattice excitations. Time-of-flight neutron diffraction shows a long-range antiferromagnetic order below $\sim 47(1)$~K, with Ru$^{5+}$ and Dy$^{3+}$ moments collinearly aligned, exhibiting nearly Ising-like anisotropy. Inelastic neutron scattering detects sharp Dy--Ru magnon modes arising from strong 4$d$--4$f$ exchange, alongside distinct Dy$^{3+}$ crystal-electric-field excitations, which a point-charge model well captures. Raman spectroscopy confirms the key phonon modes and the CEF excitation. SpinW modelling of spin wave quantifies the dominant Dy--Ru exchange interactions and weak off-diagonal anisotropy, providing microscopic insight into the origin of the ground state and the magnon spectrum, which is rare in this family of systems. Furthermore, machine-learning-based force-field calculations accurately reproduce the momentum-dependent phonon modes observed in INS. Collectively, these results demonstrate that 4$d$--4$f$ hybridisation and subtle single-ion anisotropy govern the quasiparticle dynamics in Ba$_2$DyRuO$_6$. Also, there is a significant overlap between the crystal electric field and lattice excitations in this compound. This motivates further efforts to determine conclusively whether these degrees of freedom are interacting. Non-magnetic substitution on the A sites may provide a mechanism to tune exchange interactions in the A$_2$RRuO$_6$ double perovskites. This diverse set of materials offers numerous routes for the design of quantum materials with bespoke properties.

\section{Acknowledgements}

T.B. acknowledges funding support from the Science and Engineering Research Board (SERB) (Anusandhan National Research Foundation (ANRF)), Government of India (Project No.~SRG/2022/000044). TB thanks the Science and Technology Facilities Council (STFC, UK) for providing neutron beam time on the WISH (proposal \#RB2368022, DOI: 10.5286/ISIS.E.RB2368022-1) and MARI (proposal \#RB2490022, DOI: 10.5286/ISIS.E.RB2490022-1) instruments at the ISIS Neutron and Muon Source. TB gratefully acknowledges financial support from the Department of Science and Technology, India, for access to the experimental facility at STFC and financial support to experiment with the DST-RAL project, and Jawaharlal Nehru Centre for Advanced Scientific Research (JNCASR) for managing this funding as the Nodal Office. TB thanks the Oak
Ridge National Laboratory (ORNL) for providing access
to the computational facilities for machine-learning-
based phonon calculations. Computational analysis of spin-wave dispersions using SpinW was carried out with support from the MATLAB academic license at RGIPT. M.B.S. acknowledge support from the U.S. Department of Energy (DOE), Office of Science, Basic Energy Sciences, Scientific User Facilities Division. D.T.A. acknowledges EPSRC UK (Grant Ref: EP/W00562X/1) for funding. J.S. would like to thank SERB, DST-India, for the Ramanujan Fellowship (Grant No. RJF/2019/000046).

\section*{Competing interests}
The authors declare no competing financial interests.

\end{document}

\title{Supplemental Material for\\
\textit{``Quasiparticle Dynamics in the 4d--4f Ising-like Double Perovskite Ba$_2$DyRuO$_6$ studied using Neutron Scattering and Machine-Learning Framework''}
}

\author{G. Roy}
\email{gourabr22bs@rgipt.ac.in}
\affiliation{Rajiv Gandhi Institute of Petroleum Technology, Jais, Amethi, 229304, India}

\author{E. Kushwaha}
\affiliation{Rajiv Gandhi Institute of Petroleum Technology, Jais, Amethi, 229304, India}

\author{M. Kumar}
\affiliation{Rajiv Gandhi Institute of Petroleum Technology, Jais, Amethi, 229304, India}

\author{S. Ghosh}
\affiliation{Rajiv Gandhi Institute of Petroleum Technology, Jais, Amethi, 229304, India}

\author{F. Orlandi}
\affiliation{ISIS Neutron and Muon Source, STFC, Rutherford Appleton Laboratory, Harwell campus, Didcot, Oxfordshire OX11-0QX, United Kingdom}

\author{M. D. Le}
\affiliation{ISIS Neutron and Muon Source, STFC, Rutherford Appleton Laboratory, Harwell campus, Didcot, Oxfordshire OX11-0QX, United Kingdom}

\author{M. B. Stone}
\affiliation{Neutron Scattering Division, Oak Ridge National Laboratory, Oak Ridge, Tennessee 37831, USA}

\author{J. Sannigrahi}
\affiliation{School of Physical Sciences, Indian Institute of Technology Goa, Farmagudi, Goa 403401, India}

\author{D. T. Adroja}
\affiliation{ISIS Neutron and Muon Source, STFC, Rutherford Appleton Laboratory, Harwell campus, Didcot, Oxfordshire OX11-0QX, United Kingdom}
\affiliation{Highly Correlated Matter Research Group, Physics Department,University of Johannesburg, Auckland Park 2006, South Africa}

\author{T. Basu}
\email{tathamay.basu@rgipt.ac.in}
\affiliation{Rajiv Gandhi Institute of Petroleum Technology, Jais, Amethi, 229304, India}

\maketitle

\setcounter{section}{0}
\renewcommand{\thesection}{S\arabic{section}}
\renewcommand{\thefigure}{S\arabic{figure}}
\renewcommand{\thetable}{S\arabic{table}}
\renewcommand{\theequation}{S\arabic{equation}}

\section{ Neutron Powder Diffraction}

 The Miller indices corresponding to all the $17$ magnetic peaks are listed in TABLE~S1.

\begin{table*}
\caption{Observed magnetic Bragg peaks with corresponding momentum transfer $Q$ and indexed Miller indices $(hkl)$.}
\label{tab:mag_peaks}
\begin{ruledtabular}
\begin{tabular}{|c|c|c|}
Magnetic Peak No. & $Q$ (\AA$^{-1}$) & Miller indices $(hkl)$ \\
\hline
1  & 0.75 & (000) \\
2  & 1.06 & ($\bar{1}\bar{1}1$), ($1\bar{1}1$) \\
3  & 1.68 & ($0\bar{2}2$), (002), ($2\bar{2}0$), (200) \\
4  & 1.84 & ($\bar{1}\bar{3}1$), ($\bar{1}11$), ($1\bar{3}1$), (111) \\
5  & 2.24 & ($\bar{2}\bar{2}2$), ($\bar{2}02$), (020), ($2\bar{2}2$), (202) \\
6  & 2.38 & ($\bar{3}\bar{1}1$), ($\bar{1}\bar{1}3$), ($1\bar{1}3$), ($3\bar{1}1$) \\
7  & 2.72 & ($0\bar{4}2$), (022), ($2\bar{4}0$), (220) \\
8  & 2.81 & ($\bar{3}\bar{3}1$), ($\bar{3}11$), ($\bar{1}\bar{3}3$), ($\bar{1}13$), ($1\bar{3}3$), (113), ($3\bar{3}1$), (311) \\
9  & 3.09 & ($0\bar{2}4$), (004), ($4\bar{2}0$), (400), ($\bar{2}\bar{4}2$), ($\bar{2}22$), ($2\bar{4}2$), (222) \\
10 & 3.19 & ($\bar{3}\bar{1}3$), ($\bar{1}\bar{5}1$), ($\bar{1}31$), ($1\bar{5}1$), (131), ($3\bar{1}3$) \\
11 & 3.44 & ($\bar{4}\bar{2}2$), ($\bar{4}02$), ($\bar{2}\bar{2}4$), ($\bar{2}04$), ($2\bar{2}4$), (204), ($4\bar{2}2$), (402) \\
12 & 3.53 & ($\bar{3}\bar{3}3$), ($\bar{3}13$), ($3\bar{3}3$), (313) \\
13 & 3.76 & (420), ($4\bar{4}0$), (040), (024), ($0\bar{4}4$) \\
14 & 3.82 & ($5\bar{1}1$), (331), ($3\bar{5}1$), (133), ($1\bar{1}5$), ($1\bar{5}3$), ($\bar{1}33$), ($\bar{1}\bar{1}5$), ($\bar{1}\bar{5}3$), ($\bar{3}31$), ($\bar{3}\bar{5}1$), ($\bar{5}\bar{1}1$) \\
15 & 4.04 & (422), ($4\bar{4}2$), (240), (224), ($2\bar{4}4$), ($2\bar{6}0$), (042), ($0\bar{6}2$), ($\bar{2}24$), ($\bar{2}\bar{4}4$), ($\bar{4}22$), ($\bar{4}\bar{4}2$) \\
16 & 4.12 & (511), ($5\bar{3}1$), ($\bar{5}11$), ($\bar{5}\bar{3}1$), (115), ($1\bar{3}5$), ($\bar{1}15$), ($\bar{1}\bar{3}5$) \\
17 & 4.62 & (151), ($1\bar{7}1$), ($\bar{1}51$), ($\bar{1}\bar{7}1$), (513), ($5\bar{3}3$), ($\bar{5}13$), ($\bar{5}\bar{3}3$), (315), ($3\bar{3}5$), ($\bar{3}15$), ($\bar{3}\bar{3}5$) \\
\end{tabular}
\end{ruledtabular}
\end{table*}

\clearpage

\section{ Inelastic Neutron Scattering}

The intensity as a function of $Q$ exhibits a clear decay in the crystal electric field (CEF) excitation region in the experimental INS data, as shown in FIGs.~S1 and S2. The CEF excitations are observed in the low-$Q$ region from 1.5 to 3~\AA$^{-1}$.  Additionally, the machine-learning-based phonon spectra indicate that phonon excitations predominantly emerge at higher momentum transfer ($Q \gtrsim 4$--5~\AA$^{-1}$) (see FIG. $15(b)$), supporting the magnetic origin of the low-$Q$ excitations.

The system exhibits long-range order with commensurate magnetic structure. Furthermore, we do not observe any asymmetry around the elastic line as shown in FIG.S$3$, which excludes any short-range magnetic interaction in this system. 

FIG.~S4 shows the excitations near 2.9, 4.9 and 9.6~meV display a wave-like variation in intensity across $Q$, forming continuous curve-like features in the $Q$--$E$ map (indicated by red and black lines). This behaviour reflects a clear energy--momentum dependence, characteristic of dispersive excitations. Moreover, FIG. S$5$ shows that the energy excitations around $3$, $5$, and $10$ meV exhibit a clear Q-dependent behaviour. This is evident from the intersection over two Q ranges, 0.4–1.8 Å-1 and 2.5–6 Å-1, where in both ranges the intensity decreases with increasing Q while maintaining peaks at ~3, 5, and 10 meV. Such Q dependence is characteristic of dispersive magnon excitations.

\begin{figure}[!ht]
    \centering
    \includegraphics[width=1\textwidth]{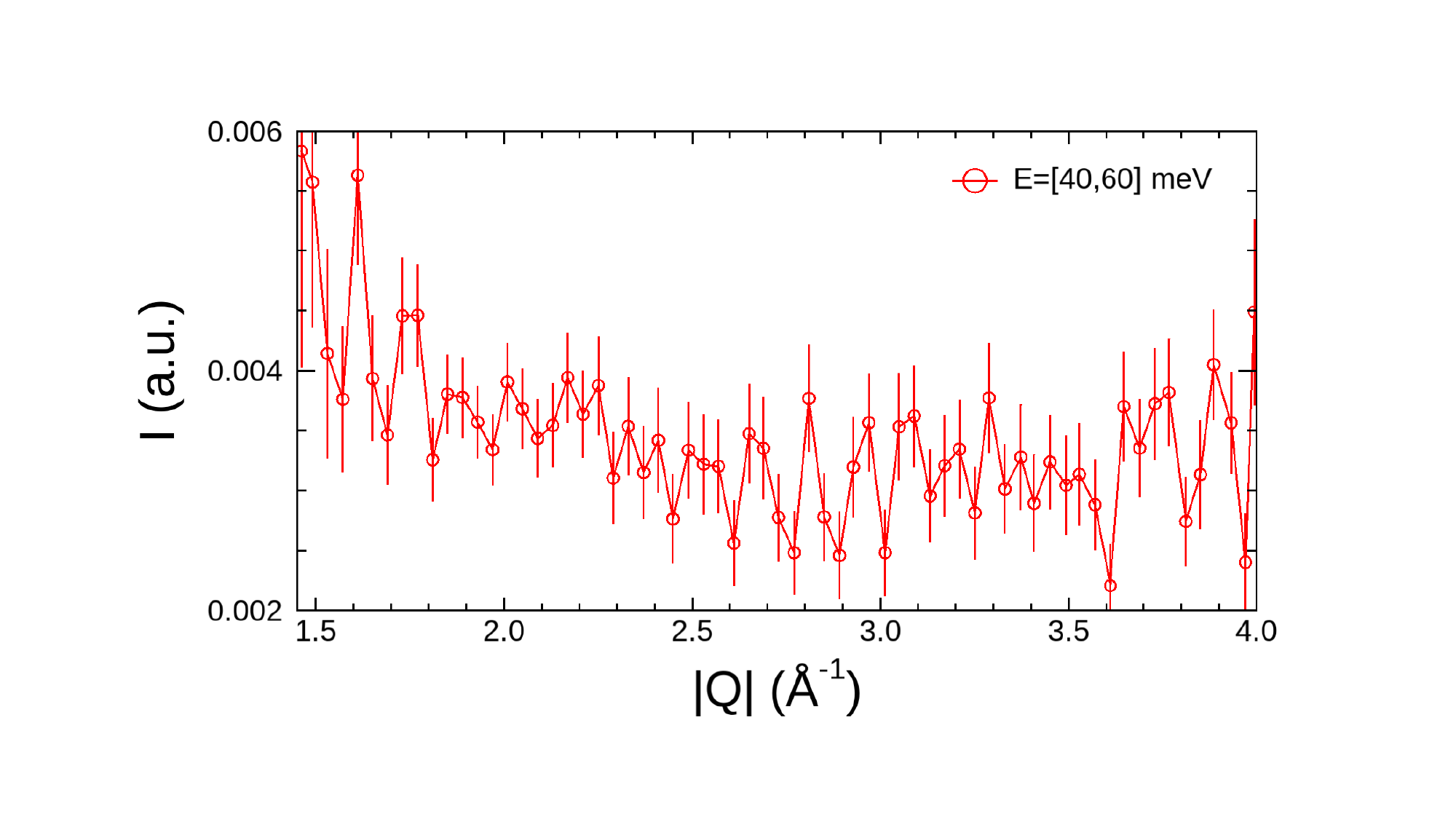} % Adjust filename and width
    \caption{Intensity as a function of momentum transfer ($Q$) for the CEF excitation around 46.5~meV, measured at 5~K using an incident neutron energy $E_i = 180$~meV.}

    \label{Fig2}
\end{figure}

\begin{figure}[!ht]
    \centering
    \includegraphics[width=1\textwidth]{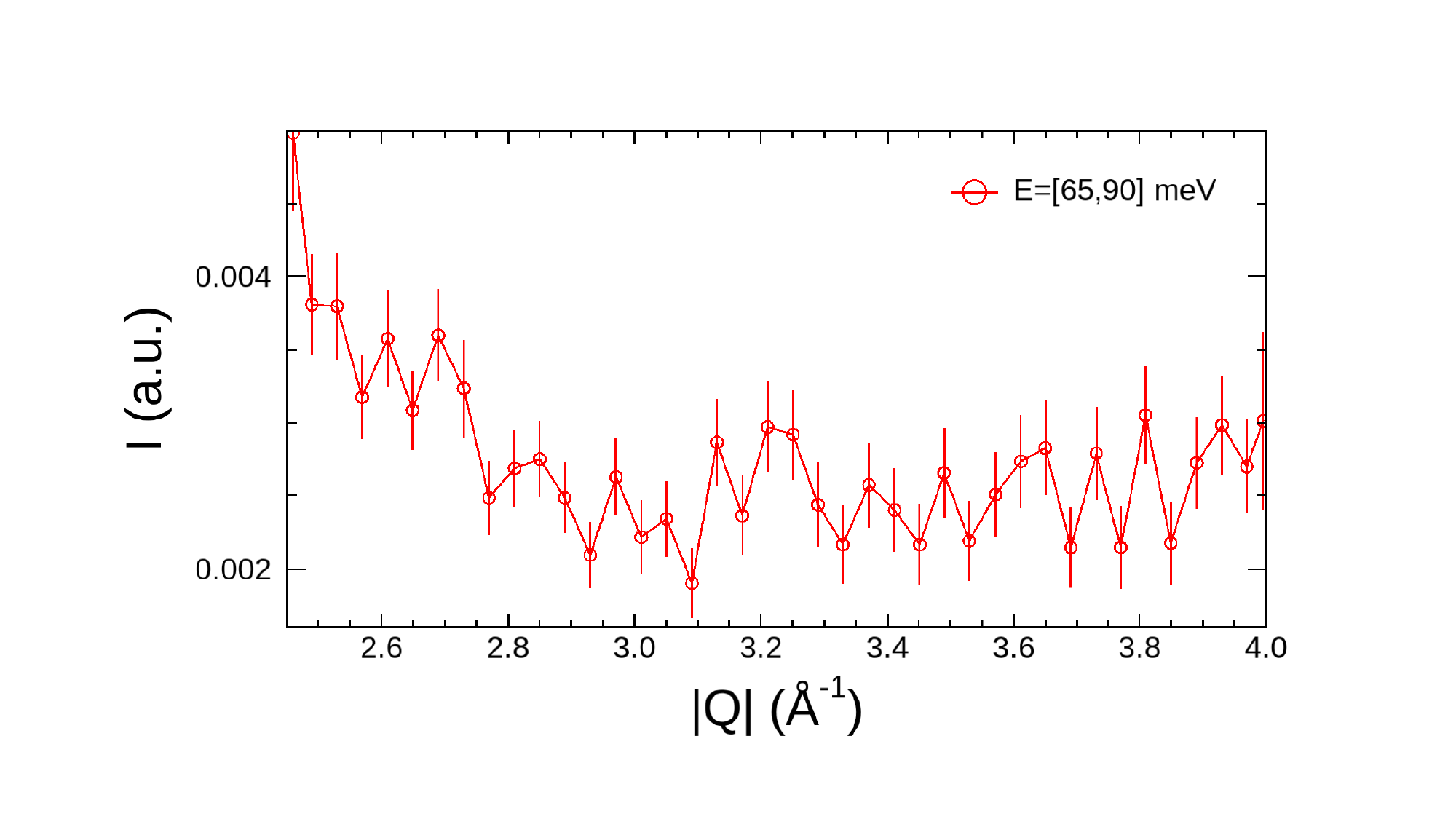} % Adjust filename and width
    \caption{Intensity as a function of momentum transfer ($Q$) for the CEF excitation around 71.8~meV, measured at 5~K using an incident neutron energy $E_i = 180$~meV.}

    \label{Fig2}
\end{figure}

\begin{figure}[!ht]
    \centering
    \includegraphics[width=1\textwidth]{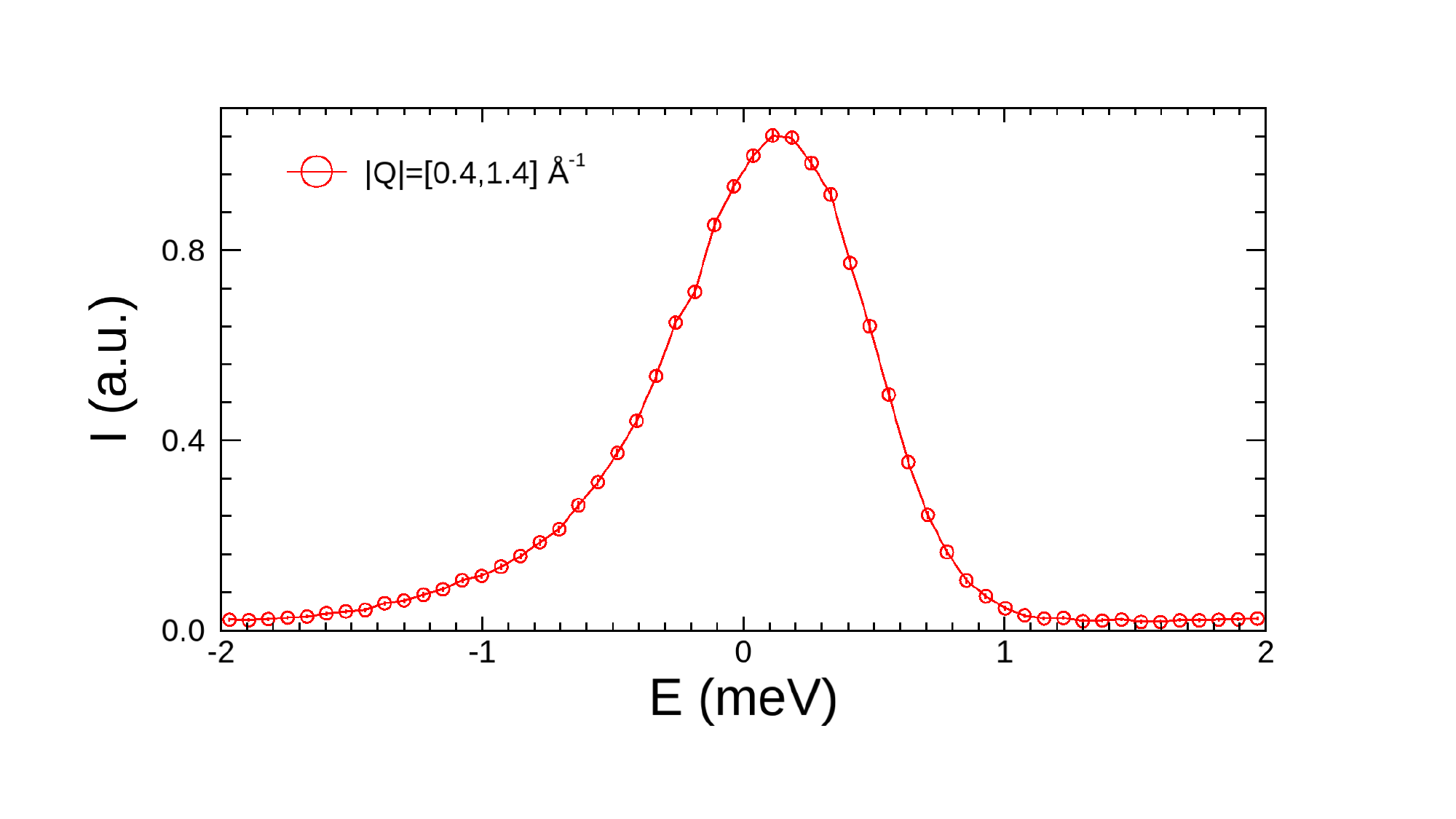} % Adjust filename and width
    \caption{Energy dependence of the neutron intensity showing symmetric behaviour about the elastic line.}

    \label{Fig2}
\end{figure}

\begin{figure}[!ht]
    \centering
    \includegraphics[width=.9\textwidth]{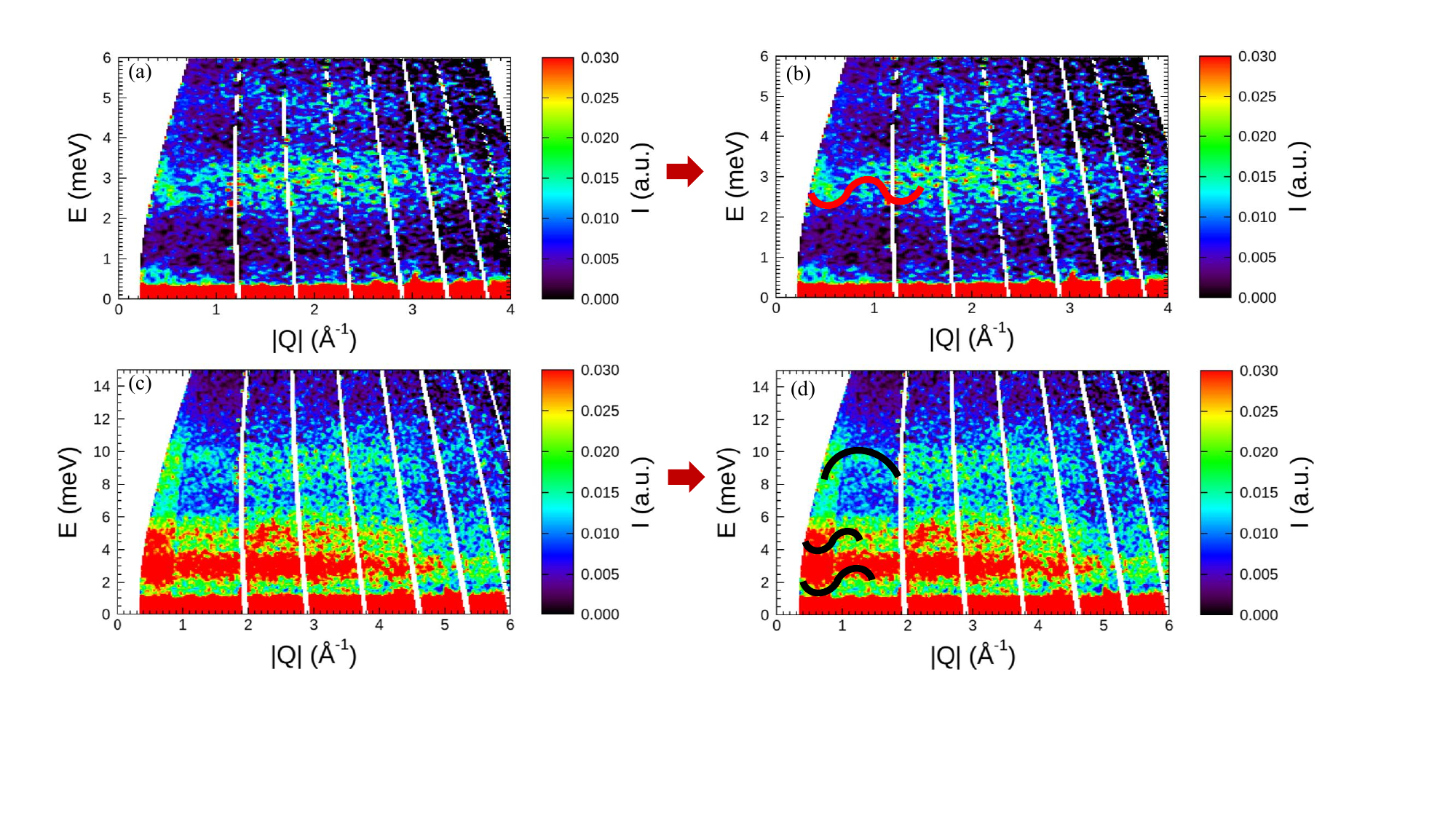} % Adjust filename and width
    \caption{Color contour plots of neutron scattering intensity in $|Q|$–$E$ space. Panels (a) and (b) show data for $E_i = 11.7$~meV, while panels (c) and (d) show data for $E_i = 29.7$~meV, revealing a wave-like dispersive nature of the magnetic excitations.}

    \label{Fig2}
\end{figure}

\begin{figure}[!ht]
    \centering
    \includegraphics[width=1\textwidth]{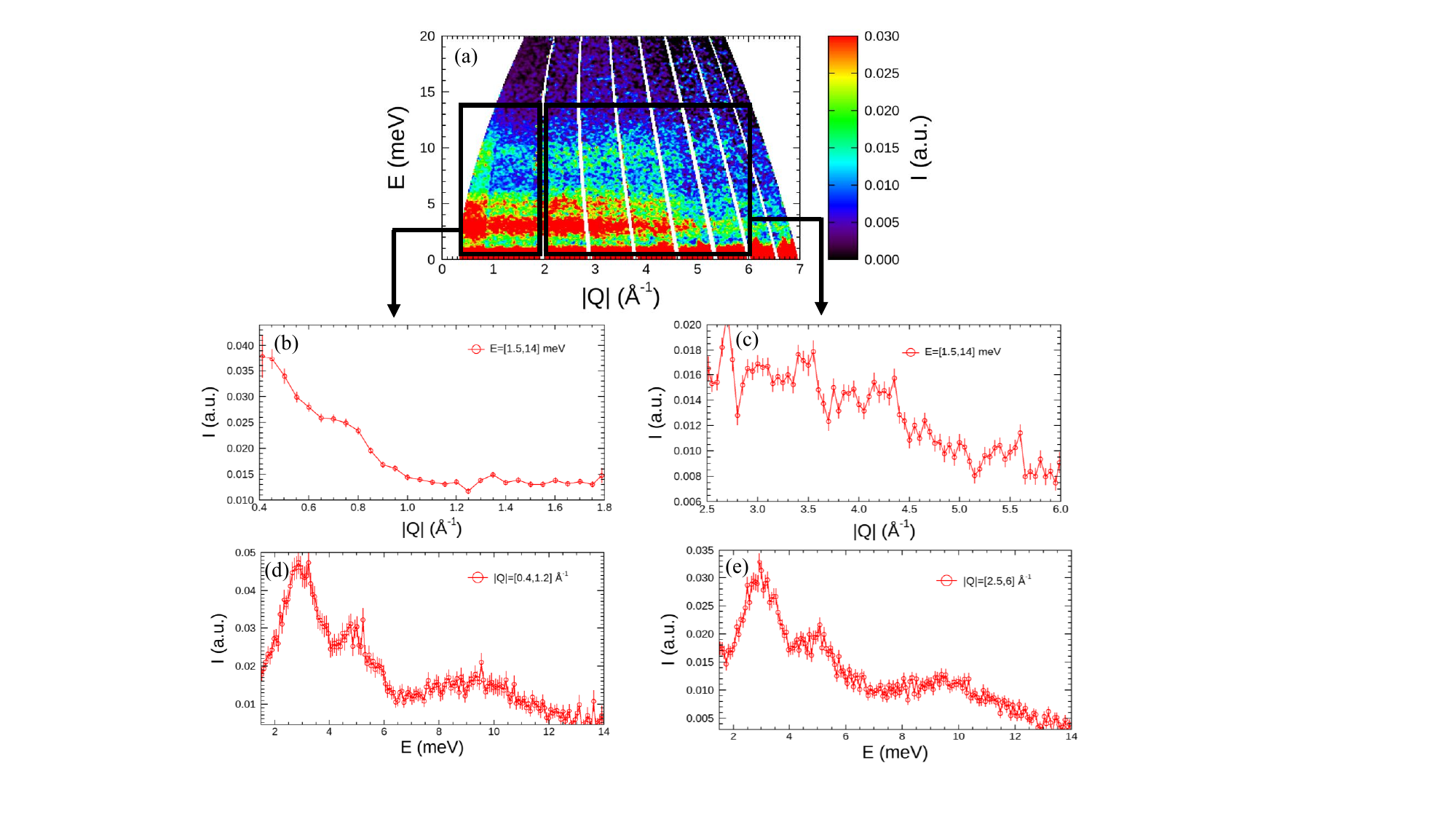} % Adjust filename and width
    \caption{Colour contour plot of neutron scattering intensity for $E_i = 29.7$~meV shown in panel (a), illustrating the excitation spectrum. Panels (b) and (c) present the $Q$-dependence of the intensity integrated over the ranges $0.4$--$1.8~\mathrm{\AA}^{-1}$ and $2.5$--$6~\mathrm{\AA}^{-1}$, respectively. Panels (d) and (e) show the energy dependence of the intensity integrated over the corresponding $Q$ ranges. }

    \label{Fig2}
\end{figure}